\documentclass[useAMS,usenatbib]{mnras}

\usepackage{amsmath}
\usepackage{amsfonts}
\usepackage{amssymb}
\usepackage{graphicx}
\graphicspath{{../Figures/}}

\usepackage[normalem]{ulem}
\usepackage{RefJ}

\usepackage{hyperref}
\usepackage[]{natbib}

\newcommand{\dd}{\mathrm{d}}

\usepackage{booktabs} 

\let\vec\boldsymbol 

\usepackage[dvipsnames]{xcolor} 

\usepackage{array}
\newcolumntype{T}{>{\tiny}l} 
\newcolumntype{H}{>{\Huge}l} 

\title[Magnetic fields from Multiplicative Chaos]{Magnetic fields from Multiplicative Chaos}
\author[J.-B. Durrive, P. Lesaffre, and K. Ferri\`ere]{Jean-Baptiste Durrive$^{1}$\thanks{E-mail:jdurrive@irap.omp.eu}, Pierre Lesaffre$^{2}$, and Katia Ferri\`ere$^{1}$\\
$^{1}$ Institut de recherche en astrophysique et plan\'etologie – Universit\'e Toulouse III - Paul Sabatier, Observatoire Midi-Pyr\'en\'ees,\\
Centre National de la Recherche Scientifique, UMR5277 – France\\
$^2$ Laboratoire de physique de l'\'ENS - ENS Paris – Centre National de la Recherche Scientifique, \'Ecole normale sup\'erieure - Paris :\\
FR684, Université Paris Diderot - Paris 7, Sorbonne Universite, UMR8023 – France}

\begin{document}

\date{Accepted 2020 May 22 ; Received 2020 May 22 ; in original form 2020 January 6}

\pagerange{\pageref{firstpage}--\pageref{lastpage}} \pubyear{2020}

\maketitle

\label{firstpage}
	
\begin{abstract}
An analytical model for fully developed three-dimensional incompressible turbulence was recently proposed in the hydrodynamics community, based on the concept of multiplicative chaos. It consists of a random field represented by means of a stochastic integral, which, with only a few parameters, shares many properties with experimental and numerical turbulence, including in particular energy transfer through scales (the cascade) and intermittency (non-Gaussianity) which is most conveniently controlled with a single parameter. Here, we propose three models extending this approach to MHD turbulence. Our formulae provide physically motivated 3D models of a turbulent velocity field and magnetic field coupled together. Besides its theoretical value, this work is meant to provide a tool for observers: a dozen of physically meaningful free parameters enter the description, which is useful to characterize astrophysical data.
\end{abstract}


\section{Introduction}

Magnetic fields are ubiquitous in the Universe, with a stunningly wide range of strengths and coherence lengths, from $10^{15}$G in magnetars to $10^{-6}$G in galaxy clusters, or even weaker at even larger scales. Indeed, while historically astrophysical plasmas pertained to solar and stellar physics, magnetism nowadays is also the focus of physicists studying much larger scales. For example, a multitude of high-resolution observations indicate that the interstellar medium of our galaxy is a magnetized, multi-phase, highly turbulent and intermittent medium \citep[e.g.][]{FalgaroneEtAl91,MAMDEtAl03,ElmegreenScalo04,Planck16_GalacticDust}. Extra-galactic fields, up to cosmological scales (Mpc size filaments, and, as claimed by an increasing number of groups, inside cosmic voids) are about to be routinely observed as well, thanks for example to the radio-telescopes LOFAR and the SKA. For reviews focused on such fields, see for instance \cite{Widrow02,RyuEtAl12,Subramanian19}. The bottom line is that, in order to understand the dynamics of astrophysical media, one needs to be able to model MHD turbulence. For astrophysically-oriented reviews on MHD turbulence see e.g. \cite{SchekochihinCowley07,BrandenburgLazarian13,Ferriere19,Tobias19}.

Astrophysical systems are fascinating for the same reason that they are so difficult to analyze: what we observe is the result of extremely rich dynamics, consisting of the intertwining of numerous processes governed by a variety of timescales, operating simultaneously at interdependent lengthscales. For this reason, solving consistently all the equations relevant for the interstellar medium for instance is not reachable analytically, and the relevant tools for this challenge are numerical simulations. Here our approach is to derive analytical expressions which, while capturing as much of the underlying physics as possible, remain simple enough to be discussed intuitively. For example, in our model intermittency is controlled by a single parameter, while in principle it is the result of complex dynamics, namely turbulent dynamo action and random shock compression. Hence, the description introduced here is phenomenological, and may be seen as a complementary tool to numerical simulations.

This work was carried out with the intention of providing a tool to describe effectively observations, grasping a fair amount of physics with only a dozen of physically meaningful free parameters. Indeed, our formulae are random processes that contain parameters (and functions) corresponding to: integral scales (at which energy is injected), dissipative scales, the large-scale shape of some mean field, the nature of the process (positively or negatively correlated increments), energy cascades (transfer through scales) and intermittency. Another difficult property to take into account in modeling MHD turbulence, is the coupling of velocity and magnetic fields. Our formulae do account for this in a realistic manner, to a certain extent, as they are built from equations embodying the dynamics.
To be able to model such features opens up many applications for astrophysics. For instance, in the interstellar medium the propagation of cosmic rays is very sensitive to the intermittency of the magnetic field in which they move \citep{ShukurovEtAl17}. Nowadays, to explore the correlation between magnetic fields and cosmic rays, one has to run costly numerical simulations \citep{SetaEtAl18}, while with the present model synthetic data may be obtained orders of magnitude faster and controlling the degree of intermittency with a single parameter.

In addition to being a vast subject, astrophysics inherits the intrinsic complexity of the physics itself. Indeed, it is well-known that turbulence per se, even in the incompressible hydrodynamical case, brings serious theoretical challenges, leaving us today with many long-lasting open questions. Exact solutions of the Navier-Stokes equations were found only in quite limited cases, and only a few statistical laws were derived rigorously. Meanwhile, various phenomenologies for MHD turbulence have been suggested in the literature. The study of Alfv\'enic turbulence has a long history, with pioneering work by \cite{Iroshnikov63} and \cite{Kraichnan65} who generalized Kolmogorov's isotropic model. A quantitative picture of the properties of anisotropic cascade emerged with the seminal papers by \cite{GoldreichSridhar95_Part2,NgBhattacharjee96}, at the heart of which lies the concept of critical balance. This picture keeps being refined, with additional ingredients such as dynamical alignment \citep{Boldyrev06,MalletSchekochihin17} and thanks to important on-going efforts to tackle the problem by means of numerical simulations; for reviews see for example \cite{Verma04,Pouquet15,Rincon19,MoffattDormy19}.
Our work relies on results derived by members of the hydrodynamics community, who themselves adapted ideas developed in the mathematics community. The latter physicists also consolidated their formal results by benefiting from close interactions with laboratory experiments. To the best of our knowledge, their approach has not been extended to MHD\footnote{The closest to our work we found is \cite{HaterEtAl11}, who also have as a starting point ideas from L.Chevillard and his collaborators, but they generalize to magnetic fields other aspects than we do, namely the time evolution of Lagrangian fluctuations.}, which is the purpose of the present paper, and has not been mentioned in the astrophysics literature yet either.

More specifically, in a series of papers \citep[essentially][]{RobertVargas08,ChevillardEtAl10,ChevillardHDR,PereiraEtAl16} a method was proposed to construct analytical expressions of 3D random vectorial fields representing the local structure of homogeneous, isotropic, stationary (fully developed) and incompressible turbulence. The random nature of turbulence is dealt with by means of stochastic calculus, i.e. employing a statistical description of turbulence, modeling fluids in terms of random functions \citep{Panchev72,MoninYaglom75}. The gist of the method is the following. Starting from a vectorial Gaussian random field, features characteristic of turbulent flows are included one by one, through a series of cunning modifications. Non-Gaussianity, i.e. intermittency, is brought in the model through the (pleasantly named) Gaussian `multiplicative chaos' introduced by \cite{Kahane85}. Applications of multiplicative chaos go well beyond the theory of turbulence, as it is used not only in mathematics, but also in finance \citep{DuchonEtAl12} or quantum gravity \citep{KnizhnikEtAl88,KupiainenEtAl18} for example \citep[see][for a recent review]{RhodesVargas14}. 
Then to obtain energy cascades, the key is to extend the scalar theory of \cite{Kahane85} to matrices. This was first done by \cite{ChevillardEtAl13} in view of applications to turbulence. In fact, one of the important successes of the phenomenology outlined here is that this velocity field is the first stochastic process proposed in the literature that is able to predict a non-vanishing mean energy transfer across scales \citep{PereiraEtAl16}. The long range correlated nature of turbulence is incorporated heuristically using a non linear transformation inspired by the `recent fluid deformation approximation' proposed by \cite{ChevillardMeneveau06}. Finally, note that in the aforementioned literature, both the Lagrangian and Eulerian descriptions are adopted. Here we stick to the Eulerian framework. 

The spirit and organization of the paper is the following. In the line of \cite{ChevillardEtAl10}, our approach is practical in the sense that we reduce mathematically involved material to intuitive considerations, such that hopefully the reader will feel that the model is built as blocks assembled gradually through a series of intuitive steps. To this end, we begin with the simple case of a scalar field in section \ref{section:ScalarField}, which introduces basic ideas. In section \ref{section:VelocityField}, the more involved case of a velocity field is treated, adapting previous works to our purpose. This allows us to prepare the ground for our generalization, makes the paper self-consistent, and advocates for digging in this direction. Indeed, at the end of this section, we emphasize how well the results of this procedure agree with direct numerical simulations and experimental data. Next, in section \ref{section:MHD}, we present our extension to magnetic fields which should appear natural, having carefully detailed the hydrodynamical case. In section \ref{section:Examples}, we illustrate some properties of the formulae of the paper through numerical examples. Finally, we conclude by giving hints of the prospects of this work.

General comments on the notations: (i) To lighten the paper, we use everywhere the shorthand notation
\begin{equation}
\vec{r} \equiv \vec{x}-\vec{y} \hspace{0.5cm} \text{and} \hspace{0.5cm} r \equiv |\vec{r}|,
\label{shortHandNotations}
\end{equation}
as most expressions are convolutions, with $r$ corresponding to the distance from a given position $\vec{x}$ when integrating with respect to $\vec{y}$. (ii) We use subscripts $s,v$ and $m$ for quantities associated with scalar fields, velocity fields and magnetic fields respectively.
\textcolor{Black}{(iii) The symbol tilde $(\sim)$ evoques fluctuations, so we use tildes to indicate random variables (modeling turbulent fields), and we add a subscript $g$ to random fields with Gaussian statistics. For example, adding a noise term in the expression of a smooth non-random field $f(\vec{x})$ results in a random field that we denote $\widetilde{f}(\vec{x})$ if it is intermittent, and $\widetilde{f}_g(\vec{x})$ if it is not.}

\section{Scalar field}
\label{section:ScalarField}

Many scalar fields in nature (e.g. density or energy dissipation) are observed to be intermittent, so it is important to find an efficient way of modeling intermittent scalar fields. To do so, let us start with a formula that we are all familiar with : the gravitational potential $\phi$ due to a given distribution of mass $\rho$ is given by
\begin{equation}
\phi(\vec{x}) = -G \int_{\mathbb{R}^3} \frac{\rho}{r} \dd V,
\label{gravitationalPotential_rho}
\end{equation}
where $G$ is Newton's constant. We may reformulate this in terms of the gravitational field $\vec{g}$, using Poisson's equation $\vec{\nabla} \cdot \vec{g} = - 4 \pi G \rho$ and the divergence theorem (assuming that the functions vanish sufficiently fast to infinity for surface integrals to be discarded), namely
\begin{equation}
\phi(\vec{x}) = \frac{1}{4 \pi} \int_{\mathbb{R}^3} \frac{\vec{r} \cdot \vec{g}}{r^3} \dd V.
\label{gravitationalPotential}
\end{equation}
Since $\vec{g} = - \vec{\nabla} \phi$ the integral operator in \eqref{gravitationalPotential} acting on $\vec{g}$ may be seen as the inverse of the gradient operator, and similarly since $\Delta \phi = 4 \pi G \rho$ the integral operator in \eqref{gravitationalPotential_rho} acting on $\rho$ may be seen as the inverse of the Laplacian.

Expression \eqref{gravitationalPotential} is an example of a physically relevant scalar field. Now, in order to construct a larger class of such scalar fields, we are going to apply several substitutions to this expression, while keeping its overall structure. First, rather than integrating up to infinite distances, let us consider an arbitrary boundary in the integral, i.e. in \eqref{gravitationalPotential} make the substitution
\begin{equation}
\mathbb{R}^3 \rightarrow \mathcal{R}_s,
\label{substitution_integrationDomain_sf}
\end{equation}
where $\mathcal{R}_s$ will now denote the region of integration. Second, let us generalize the power of the power-law kernel
\begin{equation}
r^{-3} \rightarrow r^{-2 h_s},
\end{equation}
where $h_s$ is a free parameter. Third, to avoid the singularity at the origin $r=0$ in the integrand brought by this power-law behavior, let us make the substitution
\begin{equation}
r \rightarrow (r^2+\epsilon_s^2)^{1/2},
\label{substitution_regularization}
\end{equation}
where $\epsilon_s$ is another free parameter. This is called a \textit{regularization}. Mathematically it will spare us difficulties due to divergences and ensure differentiability, while physically the parameter $\epsilon_s$ will efficiently model dissipative effects at small scales, hence this choice of notation traditionally used for small parameters.
\textcolor{Black}{The final substitution, which we call \textit{randomization}, consists of replacing the vector field $\vec{g}$ by a random field
\begin{equation}
\vec{g} \rightarrow \widetilde{\vec{\eta}}_g,
\end{equation}
where the components of the vector $\widetilde{\vec{\eta}}_g$ are Gaussian white noises (the subscript $g$ stands for `Gaussian'), independent of one another, zero-averaged, and with unit variance. This substitution modifies the nature of the integral \eqref{gravitationalPotential}, as it is now a stochastic integral. We give more information on this point and related technical aspects in appendix \ref{appendix:TechnicalAspects}.
}

The above substitutions transform the integral \eqref{gravitationalPotential} into the following stochastic representation of a scalar random field
\begin{equation}
\widetilde{s}_g(\vec{x}) = \int_{\mathcal{R}_s} \frac{\widetilde{\vec{\eta}}_g \cdot \vec{r}}{(r^2+\epsilon_s^2)^{h_s}} \dd V.
\label{s_gaussian}
\end{equation}
\textcolor{Black}{Beware, $\widetilde{s}_g$ is a general scalar field: it does not bear the physical meaning of a gravitational potential and $\widetilde{\vec{\eta}}_g$ in \eqref{s_gaussian} is just some Gaussian white noise vector, not a gravitational acceleration field. Indeed, we started this section mentioning gravity only as an illustration, firstly to build an intuition for where a shape such as \eqref{s_gaussian} may come from, and secondly because this procedure of starting with a well known formula and modifying it with the above substitutions is at the heart of what follows, as we will generalize it to the vector case in section \ref{section:BiotSavart}.}

\textcolor{Black}{As indicated by the notation (tilde and subscript $g$), expression \eqref{s_gaussian} constitues a Gaussian random process,} because it is defined as a linear operation (a convolution) over a Gaussian white noise $\widetilde{\vec{\eta}}_g$. In fact, this form is basically a spatial version of the fractional Brownian motion \citep[cf.][]{MandelbrotVanNess68} in which the exponent of the power-law is called the Hurst exponent, hence our choice of notation $h$ in \eqref{s_gaussian} and in all similar integrals throughout this paper. The parameter $h_s$ controls the severeness of the singularity and consequently modifies the statistics of the resulting field, as it appears in the exponent of power-law behaving structure functions. More intuitively, while in the classical Brownian motion (also called Wiener process) the increments are independent, in fractional Brownian motion when the Hurst exponent is larger (smaller) than a critical value the increments are positively (negatively) correlated, i.e. there is an increasing (decreasing) pattern in the previous steps
\textcolor{Black}{\citep[e.g.][section 13.4]{Rodean96}}.

The fact that \eqref{s_gaussian} is a Gaussian field severely restricts its domain of relevance for modeling natural phenomena. Fortunately however, it was shown that from it one may easily construct an intermittent field, by simply applying the exponential function. Indeed, consider the scalar field
\begin{equation}
\widetilde{s}(\vec{x}) = s_0(\vec{x}) \ \! \exp \left[\tau_s \ \! \widetilde{s}_g(\vec{x})\right],
\label{sf}
\end{equation}
where $\tau_s$ is a constant and $s_0(\vec{x})$ is a non-random (i.e. a usual, smooth) scalar field. As illustrated in figures\footnote{\textcolor{Black}{In order not to interrupt the reading of the paper with large figures, we gather all the figures at the end, in appendix \ref{appendix:Figures}.}} \ref{fig_sf} and \ref{fig_pdf_sf}, $\widetilde{s}$ is intermittent, its degree of intermittency being controlled by $\tau_s$, hence called the intermittency parameter. Stated in simple words, the field \eqref{sf} is indeed intermittent because the exponential is a non linear function that increases the contrast of the field $\widetilde{s}_g$, large values becoming larger and small ones becoming smaller, making large increments more frequent than in the Gaussian case, and resulting in the non-Gaussian wings in the probability density functions of figure \ref{fig_pdf_sf}. This exponential factor then blurs locally, on small scales, the smooth field $s_0(\vec{x})$ that bears the large-scale behavior.

In a more technical wording, this manner of modeling intermittency (i.e. multifractality) by defining a lognormal process with a long range correlation structure as the exponential of a Gaussian field\footnote{Which has a logarithmic covariance, i.e. $\mathbb{E}[\widetilde{s}_g(\vec{x})\widetilde{s}_g(\vec{y})] \sim \ln(L/|\vec{x}-\vec{y}|)$, with $L$ the integral length scale and $\mathbb{E}$ the expectation value.}, was introduced in \cite{Kahane85} and is called `Gaussian multiplicative chaos'. This terminology refers to Wiener chaos, not to the notion of chaos in dynamical systems, since this theory may be seen as a multiplicative counterpart of the additive Wiener chaos theory \citep[cf][]{RhodesVargas14}. Such a construction is used for example in hydrodynamics as a stochastic representation of energy dissipation, i.e. a scalar field that shares the same statistical properties as the dissipation field (same expectation value and moments of the filtered fields at a given scale), see \cite{Kolmogorov62,Mandelbrot72,ChevillardHDR} and the Kolmogorov-Obhukov model. Physically this description corresponds to a continuous version of discrete cascades \citep[cf Richardson's picture][]{Richardson1922} referred to as log-infinitely divisible multifractal processes in the mathematically oriented literature \citep[e.g.][]{BacryMuzy03,SchmittChainais07}.

\section{Velocity field}
\label{section:VelocityField}

The take home message of the previous section is that applying the simple transformation \eqref{sf} to a Gaussian scalar field is enough to mimic a real scalar field. Let us now investigate how this may be generalized to a vector field, in order to describe efficiently fluid turbulence. In the same spirit, we will first consider a Gaussian field, and modify it with a physically motivated and intuitive, yet more involved, transformation.

\subsection{Inspiration from multiplicative chaos: Introducing matrix $\widetilde{\vec{\mathcal{D}}}_g$}
\label{section:MultiplicativeChaos}

The vectorial version of the Gaussian scalar field \eqref{s_gaussian} is obtained by simply substituting a vector product for the scalar product, i.e. symbolically
\begin{equation}
\cdot \rightarrow \times.
\end{equation}
Indeed, \cite{RobertVargas08} showed that a random vector field that is Gaussian, incompressible, homogeneous, isotropic, and singular can be expressed as the stochastic integral
\begin{equation}
\widetilde{\vec{v}}_g(\vec{x}) = \int_{\mathcal{R}_v} \frac{\widetilde{\vec{\eta}}_g \times \vec{r}}{(r^2+\epsilon_v^2)^{h_v}} \dd V,
\label{vGaussian}
\end{equation}
where $\epsilon_v$ and $h_v$ are constants, and $\mathcal{R}_v$ denotes the region of integration.
Roughly speaking, the singular power-law shape of the kernel in \eqref{vGaussian} ensures the scale invariance property (structure functions, i.e. moments of velocity increments, behaving as power-laws) that is needed to satisfy Kolmogorov's $2/3$-law. For this reason \eqref{vGaussian} is a good starting point to model a turbulent velocity field, but it is not enough per se because it lacks two important features. Firstly, it is not intermittent. Secondly, it has a vanishing third-order moment of velocity increments (and actually all odd-order correlators). Therefore it does not reproduce the $4/5$-law and is not enough to model energy transfer across scales. 

\textcolor{Black}{In order to introduce intermittency, \cite{RobertVargas08} suggested disturbing the white noise vector by the measure (they work using measure theory rigorously) given by multiplicative chaos, in the spirit of \cite{Kahane85}. In other words, they replace in \eqref{vGaussian} the white noise $\widetilde{\vec{\eta}}_g$ by $\exp [\tau \ \! \widetilde{d}_g(\vec{y})+c] \ \! \widetilde{\vec{\eta}}_g$, where $\tau$ and $c$ are constants, and $\widetilde{d}_g$ is a Gaussian scalar field.} While this indeed provides an intermittent behavior, it does not yield energy transfer: the skewness of the resulting vector field vanishes at all scales. This is because the velocity increments do not exhibit asymmetry. Soon afterwards, \cite{ChevillardEtAl10} pointed out that if instead of using a scalar field ($\widetilde{d}_g(\vec{y})$ above) one uses a matrix field ($\widetilde{\vec{\mathcal{D}}}_g(\vec{y})$ below), the symmetry is broken and this limitation is bypassed: the resulting vector field contains both intermittency and an energy cascade. Based on these works, let us consider the velocity field
\begin{equation}
\widetilde{\vec{v}}(\vec{x}) = \int_{\mathcal{R}_v} \frac{[e^{\tau_\omega \widetilde{\vec{\mathcal{D}}}_g} \ \! \widetilde{\vec{\eta}}_g] \times \vec{r}}{(r^2+\epsilon_v^2)^{h_v}} \dd V,
\label{velocity_field_AvantExplicationsDhat}
\end{equation}
where $h_v$ and $\tau_\omega$ are parameters, whose values will be discussed in section \ref{section:Examples}, and $\widetilde{\vec{\mathcal{D}}}_g$ is the matrix \eqref{Dg_tilde} below. The purpose of the next section is to explain the reasons leading to this expression for $\widetilde{\vec{\mathcal{D}}}_g$.

\subsection{Inspiration from Biot-Savart's law: Linking $\widetilde{\vec{\mathcal{D}}}_g$ to vorticity $\vec{\omega}$}
\label{section:BiotSavart}

\begin{figure*}
\centering
\includegraphics[scale=0.28]{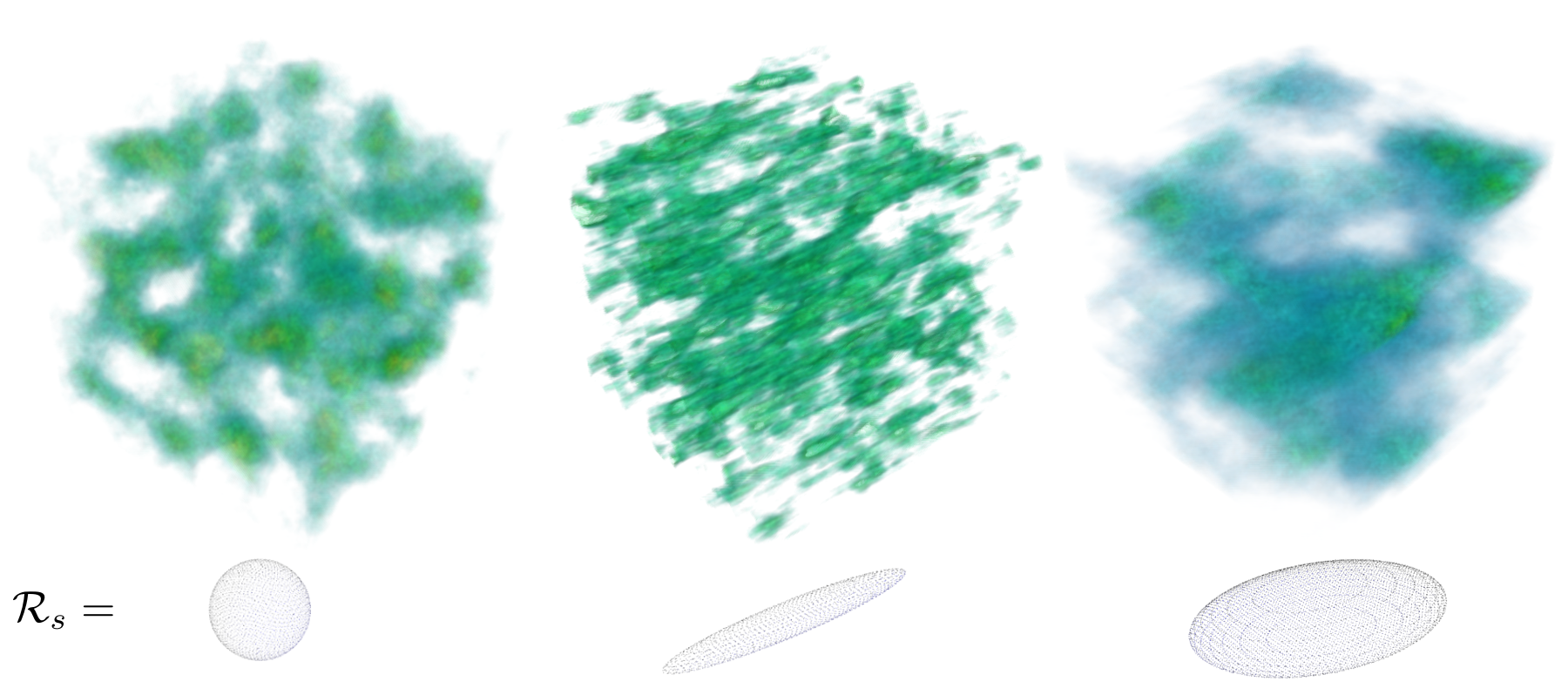}
\caption{Three scalar fields $\widetilde{s}$ built with formula \eqref{sf} from the same white noise, but varying the integration region $\mathcal{R}_s$. As illustrated in the bottom line, from left to right $\mathcal{R}_s$ has the shape of respectively a sphere, a cigare and a pancake. It appears visually that these yield respectively cloud-like, filamentary-like and sheet-like structures, matching qualitatively to the underlying shape of the region. This figure is the only part of the paper where $\mathcal{R}_s$ is not spherical. Everywhere else, we use \eqref{SphericalIntegrationRegion} i.e. spheres of radii equal to the injection scales $L_s,L_v$ of $L_m$.}
\label{fig_EllipsoidalIntegrationRegion}
\end{figure*}

Expression \eqref{velocity_field_AvantExplicationsDhat} should look familiar to the reader: it can simply be seen as a modified Biot-Savart's law. Indeed, recall that for an incompressible fluid, velocity may be expressed in terms of vorticity ($\vec{\omega} \equiv \vec{\nabla} \times \vec{v}$) according to Biot-Savart's law \citep{MajdaBertozzi01}
\begin{equation}
\vec{v}(\vec{x}) = \frac{1}{4 \pi} \int_{\mathbb{R}^3} \frac{\vec{\omega} \times \vec{r}}{r^3} \dd V,
\label{BiotSavart_Hydro}
\end{equation}
and starting with \eqref{BiotSavart_Hydro}, one obtains \eqref{velocity_field_AvantExplicationsDhat} with the four following, physically meaningful, substitutions. Firstly, the integration domain is limited
\begin{equation}
\mathbb{R}^3 \rightarrow \mathcal{R}_v.
\label{substitution_integrationDomain}
\end{equation}
Physically, this introduces correlation lengths, since the region $\mathcal{R}_v$ controls what points in space (i.e. what $\vec{y}$ in the integrand, \textcolor{Black}{because these are convolutions with $\vec{r}=\vec{x}-\vec{y}$}) influence the dynamics at position $\vec{x}$. This phenomenologically models the injection scale of the turbulent cascade operating in a given fluid state. \textcolor{Black}{If one takes an anisotropic region, several lengthscales are introduced. Points are correlated up to different distances in the various directions. Consequently, this allows one to fiddle with the geometry of the field, as illustrated in figure \ref{fig_EllipsoidalIntegrationRegion} in the scalar field case.} In section \ref{section:Examples}, we will take $\mathcal{R}_v$ to be a sphere of radius equal to the injection scale, and eddies of that size will be visible (see figures \ref{fig_vf} or \ref{fig_mfJ0} for example).
Secondly, the kernel is generalized:
\begin{equation}
r^{-3} \rightarrow r^{-2 h_v},
\label{substitution_kernelGeneralized}
\end{equation}
which allows us to model various types of correlations, as does the Hurst exponent $H$ in fractional Brownian motion. \textcolor{Black}{Indeed in fractional Brownian motion this coefficient relates to the persistence of the time series. When $H$ is greater than a half, signals have positive values that persist for a while, before switching to negative values that, in turn, also persist, and so on. On the contrary, for $H$ smaller than a half the signal `anti-persists', in the sense that it is wildly fluctuating. Here we are not dealing with time series (fractional Brownian motion) but a spatial version of it (fractional Brownian fields). Therefore the Hurst exponents in our work are not measures of long-term memory but of longe-range correlations. These coefficients enable us to control the smoothness of the fields, by modifying how they are correlated in space. For instance, \cite{ChevillardEtAl10} choose the power-law entering their rate-of-strain matrix equal to $7/2$ which imposes logarithmic long-range correlations. \textcolor{Black}{Numerical values are discussed in section \ref{section:Examples}.}} Thirdly, the norm is regularized, as was done in \eqref{substitution_regularization} in the case of a scalar field. This again mathematically guarantees that \eqref{velocity_field_AvantExplicationsDhat} is differentiable, and physically gives a degree of freedom to model dissipation at small scales (see section \ref{sec:Assets} for more information on dissipation in this model).
Finally, in order to complete this analogy, we need the following fourth substitution
\begin{equation}
\vec{\omega} \rightarrow e^{\tau_\omega \widetilde{\vec{\mathcal{D}}}_g} \ \! \widetilde{\vec{\eta}}_g,
\label{substitution_vorticity}
\end{equation}
but as such, giving a physical meaning to this is not obvious at all, especially because at this stage we do not even have an expression for $\widetilde{\vec{\mathcal{D}}}_g$. However, exhibiting the substitution \eqref{substitution_vorticity} hints towards a connection between vorticity and matrix $\widetilde{\vec{\mathcal{D}}}_g$. This is surprising a priori: $\widetilde{\vec{\mathcal{D}}}_g$ originates from the concept of multiplicative chaos, which was suggested in works unrelated to fluid turbulence. Still, led by this intuition, let us in this section have a closer look at the dynamics of vorticity. As we will see in the next section, interpreting expression \eqref{velocity_field_AvantExplicationsDhat} as stemming from \eqref{BiotSavart_Hydro} will pay off, as it will lead us to a relevant explicit expression for matrix $\widetilde{\vec{\mathcal{D}}}_g$, and bring a physical meaning to \eqref{velocity_field_AvantExplicationsDhat}.

Momentum conservation for an inviscid fluid of pressure $p$ and constant density $\rho_0$ in a gravitational field $\vec{g}$ reads
\begin{equation}
\frac{\dd \vec{v}}{\dd t} \equiv \partial_t \vec{v} + \vec{v} \cdot \vec{\nabla} \vec{v} = - \frac{1}{\rho_0} \vec{\nabla} p + \vec{g}.
\label{MomentumConservation_Hydro}
\end{equation}
For now we are working in the inviscid limit because ultimately the substitution \eqref{substitution_regularization} will take dissipative effects into account phenomenologically. The transport equation for vorticity is obtained by taking the curl of \eqref{MomentumConservation_Hydro}, recalling that $\vec{\omega}$ and $\vec{v}$ are divergence-free under the present assumptions. Doing so yields
\begin{equation}
\frac{\dd \vec{\omega}}{\dd t} = \vec{G} \cdot \vec{\omega},
\label{transportEquation1}
\end{equation}
where, to improve readability, we denote by $\vec{G}$ the gradient of velocity:
\begin{equation}
\vec{G} \equiv (\vec{\nabla} \vec{v})^\textsc{T},
\end{equation}
with $\textsc{T}$ indicating transposition.

Before going further, let us build some intuition of the underlying dynamics, by analyzing the flow locally. Taylor expanding the velocity field at a given point $\vec{x}_0$ and a given time $t$, to linear order in $|\vec{x}-\vec{x}_0|$ we have
\begin{equation}
\vec{v}(\vec{x})=\vec{v}(\vec{x}_0)+\vec{G}(\vec{x}_0)(\vec{x}-\vec{x}_0).
\label{Taylor}
\end{equation}
Next, let us decompose the matrix $\vec{G}$ into its symmetric and antisymmetric parts as
\begin{equation}
\vec{G} = \vec{\mathcal{D}} + \vec{\Omega},
\label{SymmAntisymmDecomposition_partialivj}
\end{equation}
where
\begin{equation}
\renewcommand\arraystretch{1.2} 
\begin{array}{l}
\mathcal{D}_{ij} \equiv (\partial_i v_j + \partial_j v_i)/2,\\
\Omega_{ij} \equiv (\partial_i v_j - \partial_j v_i)/2.
\end{array}
\label{Def_D_and_Omega}
\end{equation}
The symmetric part $\vec{\mathcal{D}}$ is called the deformation (or rate-of-strain) matrix. Being real and symmetric, it may be diagonalized by an orthogonal matrix, which corresponds physically to stretching and compression along its eigenvectors at a rate given by its eigenvalues. Stretching and compression occur jointly because of incompressibility. Indeed, the condition $\partial_i v_i=0$ in \eqref{Def_D_and_Omega} implies that $\vec{\mathcal{D}}$ is traceless, and thus the sum of its eigenvalues vanishes. Positive eigenvalues correspond to stretching and negatives eigenvalues to compression. The antisymmetric part $\vec{\Omega}$ is called the rotation matrix. Indeed, using the identity $\epsilon_{ijk} \epsilon_{kab} = \delta_{ia} \delta_{jb} - \delta_{ib} \delta_{ja}$, it can be written in terms of the vorticity as
\begin{equation}
\Omega_{ij} = \epsilon_{ijk} \omega_k/2,
\label{Omega}
\end{equation}
i.e. it is a cross product: In matrix form
\begin{equation}
\vec{\Omega}=\frac{1}{2} \vec{\omega}\times.
\label{Omega_matrixForm}
\end{equation}
This reveals that it corresponds to an instantaneous rotation in the direction of -$\vec{\omega}$ with angular velocity $|\vec{\omega}|/2$. As a result, the expansion \eqref{Taylor} becomes
\begin{equation}
\vec{v}(\vec{x})=\vec{v}(\vec{x}_0)+\vec{\mathcal{D}}(\vec{x}_0)(\vec{x}-\vec{x}_0)+\frac{1}{2}\vec{\omega}\times(\vec{x}-\vec{x}_0),
\end{equation}
from which we see that the flow is locally the sum of a translation, a deformation (stretching/compression) and a rotation \citep[see e.g.][]{MajdaBertozzi01}.

Introducing \eqref{SymmAntisymmDecomposition_partialivj}, together with \eqref{Omega_matrixForm}, into \eqref{transportEquation1} directly gives
\begin{equation}
\frac{\dd \vec{\omega}}{\dd t} = \vec{\mathcal{D}} \cdot \vec{\omega}.
\label{transportEquation2}
\end{equation}
Physically, this equation describes the effect of vortex stretching: Stretching a vortex along its axis reduces its cross-section due to incompressibility, and this increases its angular momentum due to conservation of angular momentum.

Hence, in order to have a closed evolution equation for $\vec{\omega}$, it remains to explicit $\vec{\mathcal{D}}$ in terms of $\vec{\omega}$. Given the definition \eqref{Def_D_and_Omega} of $\vec{\mathcal{D}}$, we need to compute the gradient of $\vec{v}$. It is tempting to simply differentiate under the integral sign in \eqref{BiotSavart_Hydro}, but unfortunately this yields a singularity and for the purpose of the present work we may simply accept the result \citep{MajdaBertozzi01}
\begin{equation}
\vec{\mathcal{D}}(t,\vec{x}) = \frac{3}{8 \pi} \mathrm{P.V.} \! \int_{\mathbb{R}^3} \! \frac{(\vec{r} \times \vec{\omega}) \ \! \vec{r} + \vec{r} \ \! (\vec{r} \times \vec{\omega})}{r^5} \dd V,
\label{VelocityGradient_SymmPart}
\end{equation}
where products like $\vec{x} \ \! \vec{y}$ are tensor products ($\vec{x} \ \! \vec{y}|_{ij}=x_i y_j$), and `P.V.' indicates a Cauchy Principal Value integral, defined as
\begin{equation}
\mathrm{P.V.} \int_{\mathbb{R}^3} f(\vec{y}) \ \! \dd V \equiv \lim_{\epsilon \rightarrow 0} \int_{|\vec{y}|\geq \epsilon} f(\vec{y}) \ \! \dd V.
\end{equation}
Equation \eqref{transportEquation2} together with \eqref{VelocityGradient_SymmPart} constitutes the equation satisfied by $\vec{\omega}$. Rather than trying to solve it exactly, \cite{ChevillardEtAl10} expose a cunning method to obtain a very efficient approximate solution, based on the short-time evolution as follows.

First, recall that equation \eqref{transportEquation2} is written in Lagrangian form, with the material derivative $\dd/\dd t = \partial_t + \vec{v} \cdot \vec{\nabla}$. Consider a Lagrangian particle at position $\vec{X}$ at an initial time $t_0$, and at position $\vec{x}(t,\vec{X})$ at a later time $t$. The vorticity $\vec{\omega}(t,\vec{X})$ felt by this particle at time $t$ satisfies \eqref{transportEquation2} where in general $\vec{\mathcal{D}}(t)$ differs from its value at $t_0$. However, after a short enough period of time, say $t-t_0 = \tau_K$ where $\tau_K$ is the Kolmogorov timescale which is the characteristic timescale of variation of the correlation of gradients, one may consider that $\vec{\mathcal{D}}(t) \approx \vec{\mathcal{D}}(t_0)$ is a good approximation. Hence, over that period of time, the matrix in equation \eqref{transportEquation2} is roughly constant, so we can approximate its solution as
\begin{equation}
\vec{\omega}(t, \vec{x}(t,\vec{X})) \approx e^{(t-t_0) \vec{\mathcal{D}}(t_0,\vec{x})} \vec{\omega}(t_0,\vec{X}).
\end{equation}
This equation shows the two processes at play: the transport of particles from position $\vec{X}$ to $\vec{x}$ (see the arguments of $\vec{\omega}$ in the right and left hand sides), and the stretching of the vorticity field by the initial deformation tensor $\vec{\mathcal{D}}(t_0)$. Now, it turns out that in the dynamics of turbulence, the components of the matrix $\tau_K \vec{\mathcal{D}}(t_0)$ are typically of order unity, while the distance traveled by particles due to advection $\tau_K v(t_0)$ is of order $\mathcal{R}_e^{-1/2}$ where $\mathcal{R}_e$ is the Reynolds number \citep{ChevillardHDR}. Besides, since in astrophysics we are mainly dealing with high Reynolds number fluids, in the following we will consider that the advection of vorticity is negligible compared to its stretching, and finally write 
\begin{equation}
\vec{\omega}(t, \vec{x}) = e^{(t-t_0) \vec{\mathcal{D}}(t_0,\vec{x})} \vec{\omega}(t_0,\vec{x}).
\label{ShortTimeDynamics_Vorticity}
\end{equation}
Since this expression stems from an expansion of equation \eqref{transportEquation2} near a given time $t_0$, in the literature this procedure is referred to as the `recent-fluid-deformation' procedure \citep{ChevillardMeneveau06,Meneveau11}.

\subsection{Putting it all together}
\label{sec:buildingDgtilde}

Let us now put the final result of section \ref{section:MultiplicativeChaos} together with that of section \ref{section:BiotSavart}: The key feature of expression \eqref{ShortTimeDynamics_Vorticity} is that this matrix exponential is reminiscent of the matrix exponential appearing in expression \eqref{velocity_field_AvantExplicationsDhat}, which completes the analogy between multiplicative chaos and vorticity. Hence, let us finally consider the velocity field
\textcolor{Black}{
\begin{equation}
\widetilde{\vec{v}} = \frac{1}{4 \pi} \int_{\mathcal{R}_v} \frac{\widetilde{\vec{\omega}} \times \vec{r}}{(r^2+\epsilon_v^2)^{h_v}} \dd V,
\label{v_tilde}
\end{equation}
where
\begin{equation}
\widetilde{\vec{\omega}} = e^{\tau_\omega \widetilde{\vec{\mathcal{D}}}_g} \widetilde{\vec{\omega}}_g,
\label{omega_tilde}
\end{equation}
and
\begin{equation}
\widetilde{\vec{\mathcal{D}}}_g = \frac{3}{8 \pi} \int_{\mathcal{R}_{\omega}} \hspace{-0.2cm} \frac{(\vec{r} \times \widetilde{\vec{\omega}}_g) \ \! \vec{r} + \vec{r} \ \! (\vec{r} \times \widetilde{\vec{\omega}}_g)}{(r^2+\epsilon_{\omega}^2)^{h_{\omega}}} \dd V.
\label{Dg_tilde}
\end{equation}
In these expressions $\widetilde{\vec{\omega}}_g$ is a Gaussian white noise vector (just like $\widetilde{\vec{\eta}}_g$ previously, but we change notation now to match its physical interpretation), 
}
$\mathcal{R}_{\omega}$ denotes the region of integration, and $h_{\omega}$ and $\epsilon_{\omega}$ are constants, with similar roles as $h_s$ and $\epsilon_s$ of the scalar case \eqref{s_gaussian}. By suggesting this form, a few additional substitutions have been made, which we now detail.

Clearly $\widetilde{\vec{\mathcal{D}}}_g$ in \eqref{Dg_tilde} stems from $\vec{\mathcal{D}}$ in \eqref{VelocityGradient_SymmPart}, with a few modifications. Firstly, we have introduced a parameter $\epsilon_{\omega}$ in the denominator, i.e. the singular kernel has been regularized, as was done in the scalar field case \eqref{substitution_regularization} and in the expression of the velocity field \eqref{v_tilde}. Thanks to this, we may discard the Cauchy Principal Value. Secondly we have generalized the power-law to some free parameter $h_{\omega}$, in order to gain an additional degree of freedom in the modeling and to have a larger parameter space for data fitting. The hydrodynamics literature we take the above formulae from is dedicated to the local structure of turbulence and ultimately aims at obtaining exact solutions of the Navier-Stokes equations. In contrast, our own purpose is different, and we do not fix a priori the parameters, e.g. we do not place constraints on the integration regions $\mathcal{R}_v$ and $\mathcal{R}_{\omega}$. This will give us more degrees of freedom, as illustrated in the scalar field case in figure \ref{fig_EllipsoidalIntegrationRegion}, to fit astrophysical data which include large-scale features. This is somewhat of a poor man's way, as a first approach, to model correlations between large and small scales, in the same spirit as we included a large-scale position dependent amplitude $s_0(\vec{x})$ in the scalar field \eqref{sf}. More refined modeling, which would for instance take into account the advection of vorticity, which is neglected in equation \eqref{ShortTimeDynamics_Vorticity}, is left for future work.
\textcolor{Black}{
Thirdly $\widetilde{\vec{\mathcal{D}}}_g$ is a `randomized' version of matrix $\vec{\mathcal{D}}$, which is what the tilde notation indicates, as we made the substitution
\begin{equation}
\vec{\omega} \rightarrow \widetilde{\vec{\omega}}_g.
\label{randomization_2}
\end{equation}
This substitution models the randomness of the stretching occurring in formula \eqref{omega_tilde} (detailed in the next paragraph) due to $\widetilde{\vec{\mathcal{D}}}_g$, since $\vec{\mathcal{D}}$ has the physical meaning of local vortex stretching. Choosing the very same white noise vector $\widetilde{\vec{\omega}}_g$ in both \eqref{omega_tilde} and \eqref{Dg_tilde} is a thoughtful decision: \cite{ChevillardEtAl10} as well as \cite{PereiraEtAl16} showed that if one considers two independent white noises, all odd-order structure functions vanish, which prevents energy transfer.} Finally, and most importantly as it is at the heart of the physical interpretation of this model, the velocity field \eqref{v_tilde} is independent of time (it describes a stationary state), which seems in contradiction with the fact that its construction is inspired from dynamics. Indeed, completing the analogy outlined in section \ref{section:BiotSavart} led us to make this last substitution
\begin{equation}
t-t_0 \rightarrow \tau_\omega,
\label{stationarization}
\end{equation}
when going from \eqref{ShortTimeDynamics_Vorticity} to \eqref{omega_tilde}. Since this amounts to substituting some parameter for the time variable characterizing the stationary state (more precisely, its intermittency), \eqref{stationarization} is referred to as \textit{stationarization}.

\textcolor{Black}{
With this in mind, the following picture emerges\footnote{Or at least let us attempt the following intuitive wording, since so far we have not found much discussion about this hypothesis in the quoted literature.}. To begin with, consider a first `timestep' (of say a Kolmogorov timescale $\tau_K$) during which expression \eqref{omega_tilde} may be seen as an evolution equation: an initial vorticity field $\widetilde{\vec{\omega}}_g$ with Gaussian statistics is stretched according to matrix $\widetilde{\vec{\mathcal{D}}}_g$, resulting in an intermittent vorticity field $\widetilde{\vec{\omega}}$. By incompressibility, through Biot-Savart's law, this translates into the velocity field \eqref{v_tilde}. What the stationarization hypothesis suggests is that this occurs at each timestep: somehow, by some undetermined non-linear process, the non-Gaussianity of the vorticity field is suddenly `blurred', resetting the underlying vorticity field to a Gaussian field $\widetilde{\vec{\omega}}_g$ again, such that, at the subsequent timestep, the stretching \eqref{omega_tilde} repeats itself. This repeated succession of stretching/blurring is such that in the end the \textit{stationary} state of the fully developed turbulent flow has similar statistics as those of a stretched Gaussian vorticity field (it does indeed, cf. the successes of this modeling detailed in section \ref{sec:Assets}). The degree of intermittency is then controlled by how long these stretching periods last, which in practice is encoded phenomenologically in the intermittency parameter $\tau_\omega$. For example if they are extremely short ($\tau_\omega \to 0$), there is not enough time for the stretching to occur significantly. The vorticity field then remains Gaussian at each timestep, resulting in a Gaussian stationary velocity field (i.e. \eqref{v_tilde} tends to \eqref{vGaussian}). The above picture also helps understand how formula \eqref{v_tilde} can describe a fully developed state of turbulence, which is the result of highly non-linear dynamics, while it was constructed from a linearization\footnote{\textcolor{Black}{Stretching itself is a non-linear process, and is what accounts for the development of non-Gaussian statistics \citep{ChevillardMeneveau06}, but considering the short-time dynamics of the transport equation \eqref{transportEquation2}, we have made its right hand side linear in $\vec{\omega}$, while in full form it is non-linear in $\vec{\omega}$.}}.
}

\textcolor{Black}{
With this interpretation, taking the very same white noise in \eqref{omega_tilde} and \eqref{Dg_tilde} becomes natural, and not only a technical requirement allowing for energy transfer: In the deformed vorticity field \eqref{ShortTimeDynamics_Vorticity} it is the initial value $\vec{\mathcal{D}}(t_0,\vec{x})$ that comes into play, so it is indeed the `initial' vorticity $\widetilde{\vec{\omega}}_g$ that makes sense in $\widetilde{\vec{\mathcal{D}}}_g$.
}

\textcolor{Black}{
A word of caution however concerning this interpretation: strictly speaking, $\widetilde{\vec{\omega}}$ is not exactly the vorticity field in this model. For example we did not impose $\widetilde{\vec{\omega}}$ to be divergence free, while this is required for a vorticity field, and because these are stochastic integrals, one should be a little bit careful with its physical dimension (see appendix \ref{appendix:TechnicalAspects}). To clarify: the procedure here is to first suggest \eqref{v_tilde} as our model for a velocity field $\widetilde{\vec{v}}$, and only then one \textit{defines} the vorticity field as the curl of $\widetilde{\vec{v}}$. All the intermediate steps outlined above should be seen more as ideas useful to construct this model, and build an intuition out of it.
}

\subsection{Assets of the adopted method}
\label{sec:Assets}

While part of the procedure presented above is based on rigorous derivations, partly taken from the mathematics literature, some points rely on phenomenological arguments. Yet, by means of both analytical and numerical arguments, previous authors have shown that the model \eqref{v_tilde} gives a surprising amount of correct predictions when confronted to both direct numerical simulations and experimental data \citep[e.g.][]{ChevillardEtAl12}. What is even more encouraging is that some of its realistic properties are not ingredients introduced by hand but by-products of this construction, which suggests that it really does capture an important part of the underlying physics. In this section we briefly review these successes.

(i) A first important feature is the existence of intermittent corrections. Let us define the velocity increment of lag $\vec{\ell}$ as
\begin{equation}
\delta_{\vec{\ell}} \vec{v}(\vec{x}) \equiv \vec{v}(\vec{x}+\vec{\ell}) - \vec{v}(\vec{x}).
\label{def:increment}
\end{equation}
Longitudinal and transverse increments are the projections of \eqref{def:increment} respectively along and perpendicular to the direction of $\vec{\ell}$.
One common way to identify intermittency is to compare the probability density functions (PDFs) of the considered field to those of a Gaussian field. And indeed the PDFs of longitudinal and transverse increments of \eqref{v_tilde} undergo a continuous deformation as the norm of the lag is decreased, from a Gaussian shape at large lags towards large tails at small lags. These often called `non-Gaussian wings' are typical signatures of intermittency. Through several examples, we show in section \ref{section:Examples} that these wings become increasingly significant as the parameter $\tau_\omega$ increases, which supports naming it the intermittency parameter. More details may be found in the above-mentioned literature.

(ii) Another classical way to characterize intermittency in isotropic turbulence studies is by analyzing the power-law behavior of structure functions in the inertial range. By definition, the $n^\text{th}$ order structure function is the $n^\text{th}$ moment of velocity increments $S_n(\ell) \equiv \langle (\delta_\ell \vec{v})^n \rangle$ where brackets $\langle \rangle$ indicate the expectation value \citep{Frisch95}. In the inertial range, $S_n \propto \ell^{\tau_n}$, where the dependence on $n$ of $\tau_n$ quantifies the intermittency: the field is intermittent if and only if $\tau_n$ depends non-linearly on $n$. For the velocity \eqref{v_tilde}, this non-linearity was confirmed numerically, and also partially analytically: under the simplifying assumption of independence of $\widetilde{\vec{\mathcal{D}}}_g$ with the white noise in \eqref{v_tilde}, \cite{PereiraEtAl16} showed that $\tau_n = a n + b n^2$, where $a$ and $b$ are constants, with $b \propto \tau_\omega^2$, i.e. the quadratic behavior is determined by the intermittency parameter. This again justifies the name given to $\tau_\omega$.

(iii) \cite{PereiraEtAl16} further characterized the non-Gaussian behavior of fluctuations by analyzing in details the skewness $\mathcal{S}_\ell \equiv \langle (\delta_\ell \vec{v})^3 \rangle/[\langle (\delta_\ell \vec{v})^2 \rangle]^{3/2}$ and flatness $\mathcal{F}_\ell \equiv \langle (\delta_\ell \vec{v})^4 \rangle/[\langle (\delta_\ell \vec{v})^2 \rangle]^2$ of velocity increments \citep{Frisch95}. The propreties of both longitudinal and transverse increments match the results of direct numerical simulations. In particular, they showed that this velocity field has a higher level of intermittency in the transverse case than in the longitudinal case, which is a feature observed in experiments \citep[see][and references therein]{PereiraEtAl16}.

(iv) The degree of freedom provided by the Hurst parameter $h_v$ allows the velocity field to satisfy the experimental $2/3$ law of turbulence.

(v) Another aspect that has been scrutinized in the literature is the statistics of the velocity gradient $\vec{G}$, which regulates the dynamics of the turbulent flow. Its eigenvalues are particularly meaningful physically (see section \ref{section:BiotSavart}) and have peculiar statistical properties that \eqref{v_tilde} reproduces remarquably well. More specifically, these eigenvalues are the solutions of the characteristic polynomial with coefficients given by the two invariants
\begin{equation}
\renewcommand\arraystretch{1.2} 
\begin{array}{l}
\displaystyle Q \equiv - \tfrac{1}{2} \text{tr}(\vec{G}^2) = \tfrac{1}{4} |\vec{\omega}|^2 - \tfrac{1}{2} \text{tr}(\vec{\mathcal{D}}^2),\\
\displaystyle R \equiv - \tfrac{1}{3} \text{tr}(\vec{G}^3) = - \tfrac{1}{4} \vec{\omega} \cdot \vec{\mathcal{D}} \cdot \vec{\omega} - \tfrac{1}{3} \text{tr}(\vec{\mathcal{D}}^3).
\end{array}
\end{equation}
Physically, $Q$ \textcolor{Black}{is a combination of enstrophy $(|\vec{\omega}|^2)$ and dissipation (both per unit viscosity) such that} when $Q>0$ rotation dominates locally in the flow, and when $Q<0$ the region is dissipation-dominated. $R$
\textcolor{Black}{is a combination of the local \textit{production} of enstrophy and dissipation, such that}
when $R>0$ the flow tends to create dissipation, and when $R<0$ more enstrophy is produced. Now, it turns out that when representing the joint probability density of $R$ and $Q$, a characteristic shape appears, known as the `teardrop shape in the RQ-plane'. See for instance \cite{Tsinober01,Wallace09,Meneveau11} for both experimental and theoretical aspects of this feature. In particular, this shape is symmetric with respect to the $R=0$ line in the Gaussian case ($\tau_\omega = 0$) and tilted anticlockwise otherwise. As detailed notably in \cite{ChevillardHDR,PereiraEtAl16}, this geometry is indeed recovered with \eqref{v_tilde}. This diagnostic tool is yet another piece of evidence of its intermittent behavior, and of the local dissipation occurring in a pretty realistic manner. 

(vi) Other essential and non-trivial features are the orientation properties of vorticity with respect to the eigenframe of the rate-of-strain matrix $\vec{D}$. In particular, the preferential alignement of vorticity with the eigenvector associated with the intermediate eigenvalue of $\vec{D}$ does not occur in the Gaussian case, but gets stronger as the intermittency parameter increases. All this is correctly reproduced by \eqref{v_tilde}, cf e.g. \cite{ChevillardHDR,PereiraEtAl16,PereiraEtAl18}.

(vii) Equation \eqref{v_tilde} is the first stochastic process proposed in the literature to have a non-vanishing mean energy transfer across scales, as observed numerically in \cite{ChevillardEtAl10} and then, by means of a perturbative expansion in the intermittency parameter, \cite{PereiraEtAl16} confirmed analytically the existence of this cascade. As this feature stems from the matrix nature of the multiplicative chaos used, for more details on mathematical aspects we refer the reader to \cite{ChevillardEtAl13} who constructed the theory of matrix multiplicative chaos.

(viii) These formulae are well behaved, notably they are differentiable thanks to the regularization. This aspect is not obvious since they are built from irregular mathematical objects (e.g. sample-paths of Wiener processes are almost nowhere differentiable).

(ix) Last but not least, another success of this approach is that it results in fields that are easy to obtain numerically. Because most equations in this work are convolutions, their implementation is both simple and fast, thanks to Fast Fourier Transform algorithms and because convolutions in Fourier space are just products. To give an order of magnitude, with only 32 CPUs one may generate a $2048^3$ resolution cube in only 5 minutes.

As a closing remark, let us add that more achievements related to this description may be found in the above-mentioned literature, notably in the Lagrangian framework, studying the time evolution. Here we stick to stationary fields in the Eulerian description. Also, here we aim at constructing expressions that are as handy as possible. For this reason, we chose the simplest regularization, namely \eqref{substitution_regularization}, but more elaborate regularized norms are used and discussed in the literature. Similarly, instead of our integration regions, it is more common to introduce large-scale cutoff functions in the kernels, which is more convenient to analyze the statistics of the field depending on the properties of these functions.

\section{Magnetic field}
\label{section:MHD}

The above substitutions constitute building blocks to be assembled to construct physically motivated turbulent vector fields. The successes mentioned in the previous section motivate us to extend this approach to magnetized fluids, which are ubiquitous in astrophysics. Per se, this is a very ambitious project, and we will limit ourselves to proposing three straightforward extensions. To do so, we need to adapt the matrix \eqref{Dg_tilde} and ensure that the magnetic fields remain divergence-free.

\subsection{Matrix field of the underlying dynamics}

Compared to the hydrodynamical case \eqref{MomentumConservation_Hydro}, the momentum equation is now coupled to the induction equation, and contains the additional contribution from the Lorentz force, $\vec{j} \times \vec{B}$, where the current density $\vec{j}$ is given by Amp\`{e}re's law $\vec{j} = \left( \vec{\nabla} \times \vec{B} \right)/\mu_0$ and the constant $\mu_0$ is vacuum permeability. In the same spirit as previously, we may omit viscosity and resistivity for now, since dissipative effects will ultimately be taken into account through regularizations similar to \eqref{substitution_regularization}. We are then left with
\begin{equation}
\renewcommand\arraystretch{1.7} 
\begin{array}{l}
\displaystyle \partial_t \vec{v} + \vec{v} \cdot \vec{\nabla} \vec{v} = - \frac{1}{\rho_0} \vec{\nabla} p + \vec{g} + \frac{1}{\mu_0 \rho_0} \left( \vec{\nabla} \times \vec{B} \right) \times \vec{B},\\
\displaystyle \partial_t \vec{B} = \vec{\nabla} \times \left(\vec{v} \times \vec{B}\right),
\end{array}
\label{IncompMHD}
\end{equation}
together with the constraint $\vec{\nabla} \cdot \vec{B} = 0$. 
As a first approach, in this paper we consider incompressible MHD and the case when the back-reaction of the magnetic field, due to the Lorentz force, is negligible. This approximation, called kinematic MHD, is valid as long as magnetic energy is small compared to kinetic energy, i.e. $B^2/(2 \mu_0) \ll 1/2 \rho_0 v^2$. This condition is often not satisfied in astrophysics. However, building our models in this framework is also motivated \footnote{\textcolor{Black}{We do not say that dynamo-produced fields will necessarily tend to be force free. What we do is build formulae taking \textit{inspiration} from force free dynamics. Then the complex intertwining of the free parameters (large-scale cut-offs of the integration regions which may be anisotropic, and the various intermittency, regularization and Hurst parameters) may make these synthetic fields mimic fields which were the result of far more complicated dynamics than the ingredients used to derive our simple expressions. This is the same idea as in the hydrodynamical case, in which the expression for $\widetilde{\vec{D}}_g$ is \textit{inspired} from the inviscid dynamics, while in the end the result impressively describes flows with dissipation (cf section \ref{sec:Assets}), thanks to the phenomenological parameter $\epsilon$.}}
by 3D MHD simulations such as those of \cite{ServidioEtAl08} which show a natural tendency towards force-free behaviors in MHD turbulence. Indeed, as nonlinearity develops, strong alignments spontaneously appear notably between $\vec{j}$ and $\vec{B}$, which is called `beltramization' in reference to Beltrami vector fields which are fields parallel to their own curl.

In this framework, the induction equation reads
\begin{equation}
\frac{\dd \vec{B}}{\dd t} = \vec{G} \cdot \vec{B}.
\label{EvolutionOfB}
\end{equation}
Introducing \eqref{SymmAntisymmDecomposition_partialivj}, together with \eqref{Omega_matrixForm}, into \eqref{EvolutionOfB} leads to
\begin{equation}
\frac{\dd \vec{B}}{\dd t} = \vec{\mathcal{D}} \cdot \vec{B} + \frac{1}{2} \vec{\omega} \times \vec{B}.
\label{EvolutionOfB_2}
\end{equation}
This equation for the magnetic field is the counterpart of equation \eqref{transportEquation2} for the vorticity. The two terms on the right hand side describe flux tube stretching (similar to vortex stretching in \eqref{transportEquation2}) and shear, respectively.
By analogy with the hydrodynamical case \textcolor{Black}{where we made the substitution \eqref{randomization_2} to construct $\widetilde{\vec{\mathcal{D}}}_g$, we suggest considering the following `randomized' velocity gradient
\begin{equation}
\widetilde{\vec{G}}_g=\widetilde{\vec{\mathcal{D}}}_g+\frac{1}{2} \widetilde{\vec{\omega}}_g \times,
\label{Ggtilde}
\end{equation}
i.e. in index notation $(\widetilde{\vec{G}}_g)_{ij}=(\widetilde{\vec{\mathcal{D}}}_g)_{ij}+\epsilon_{ijk} (\widetilde{\vec{\omega}}_g)_k/2$. Expression \eqref{Ggtilde} is the most natural choice to remain consistent with the interpretation outlined in section \ref{sec:buildingDgtilde}: it is the `initial vorticity' $\widetilde{\vec{\omega}}_g$ that comes into play in the `short-time' deformation matrix $\widetilde{\vec{G}}_g$. Also, with this choice $\widetilde{\vec{G}}_g$ is a Gaussian process, being built from a linear operation on a Gaussian process, and the purpose of the theory of multiplicative chaos is precisely to give a meaning to the exponential of such a field \citep{RhodesVargas14}.}

\subsection{Three models}

Assuming that the matrix field $\widetilde{\vec{G}}_g$ in \eqref{Ggtilde} is an accurate representation of the local dynamics in the magnetized turbulent flow, let us now propose three models for MHD turbulence, that are in principle complementary.\\

\textit{Model 1)}
By analogy with the short time expression of vorticity \eqref{ShortTimeDynamics_Vorticity} from the hydrodynamical case, it is natural to consider the short time solution of \eqref{EvolutionOfB}, namely the matrix exponential of $\widetilde{\vec{G}}_g$. Doing so, the initial value of the magnetic field appears, just like $\vec{\omega}(t_0,\vec{x})$ in \eqref{ShortTimeDynamics_Vorticity}. When suggesting the velocity field expression \eqref{v_tilde}, the initial value of vorticity was replaced by a Gaussian field through the substitution \eqref{substitution_vorticity}. Here instead, we take advantage of this degree of freedom to introduce some ordered magnetic field $\vec{B}_0$. Introducing this is important to model astrophysical fluids since the observed magnetic fields in the Universe are usually turbulent with large-scale mean fields \citep[e.g.][]{Ferriere07,Jaffe19}.
\textcolor{Black}{Hence, our first model is
\begin{equation}
\widetilde{\vec{B}}^{(1)} = e^{\tau_m \widetilde{\vec{G}}_g} \vec{B}_0,
\label{Bmodel1}
\end{equation}
where $\tau_m$ is a magnetic intermittency parameter.} Physically, the magnetic field $\widetilde{\vec{B}}^{(1)}$ is the result of the random stretching/compression and shearing (encoded in the matrix exponential, the strength of which is controlled by $\tau_m$) of a large-scale ordered magnetic field $\vec{B}_0(\vec{x})$, due to the turbulent velocity field \eqref{v_tilde} that it is coupled to. \textcolor{Black}{The divergence-free condition will be discussed at the end of this section.}\\

\textit{Model 2)}
By analogy with Biot-Savart's law \eqref{BiotSavart_Hydro} from the hydrodynamical case, it is natural to consider Biot-Savart's law for magnetic fields
\begin{equation}
\vec{B} = \frac{\mu_0}{4 \pi} \int_{\mathbb{R}^3} \frac{\vec{j} \times \vec{r}}{r^3} \dd V.
\label{BiotSavart_Magnetic}
\end{equation}
To strictly follow the procedure performed with $\vec{\omega}$, the next step would be to analyze the evolution equation of $\vec{j}$, and its short time dynamics would reveal a relevant matrix. However, here we propose a simpler and faster approach. We already found a seemingly relevant generalization of $\widetilde{\vec{\mathcal{D}}}_g$, namely $\widetilde{\vec{G}}_g$, costlessly obtained from the governing equations \eqref{IncompMHD}. And fortunately, the aforementioned beltramization exhibited in simulations of \cite{ServidioEtAl08} prompts us to indeed use $\widetilde{\vec{G}}_g$: since $\vec{j}$ tends to be aligned with $\vec{B}$, the matrix governing its evolution should in some way resemble the matrix governing the evolution of $\vec{B}$, namely \eqref{EvolutionOfB}. Hence, rather than digging further to look for the precise matrix out of the evolution of $\vec{j}$, we will directly work with $\widetilde{\vec{G}}_g$. Admittedly, this is also very convenient.
In addition, the simulations of \cite{ServidioEtAl08} also indicate that the alignments are directly related to the self-organization of the magnetofluid giving rise to spatial intermittency, which our formulae do take into account through the intermittency parameters $\tau_\omega$ and $\tau_m$. \cite{ServidioEtAl08} furthermore suggest that these dynamically generated correlations suppress nonlinearity, i.e. the equations tend to be quasilinear rather than fully nonlinear. This is reminiscent of the interpretation mentioned in section \ref{sec:buildingDgtilde} to account for the intriguing fact that our formulae \textcolor{Black}{are obtained after a linearization} to describe a fully developed turbulent state.

From the above considerations, we propose as a second model
\begin{equation}
\widetilde{\vec{B}}^{(2)} = \frac{\mu_0}{4 \pi} \int_{\mathcal{R}_m} \frac{\widetilde{\vec{j}}^{(2)} \times \vec{r}}{(r^2+\epsilon_m^2)^{h_m}} \dd V,
\label{Bmodel2}
\end{equation}
where $\mathcal{R}_m$, $h_m$ and $\epsilon_m$ have similar physical meanings as $\mathcal{R}_v$, $h_v$ and $\epsilon_v$ in the velocity field \eqref{v_tilde} (for examples see section \ref{section:Examples}). \textcolor{Black}{The field $\widetilde{\vec{j}}^{(2)}$ acts like a random current density, which, by analogy with \eqref{omega_tilde}, we build as
\begin{equation}
\widetilde{\vec{j}}^{(2)} = e^{\tau_m \widetilde{\vec{G}}_g} \ \! \widetilde{\vec{j}}_g,
\label{j_2}
\end{equation}
where $\widetilde{\vec{j}}_g$, the `undeformed' current density, is a Gaussian white noise. Note that $\widetilde{\vec{j}}_g$ must be different from $\widetilde{\vec{\omega}}_g$, otherwise, like in the hydrodynamical case, the antisymmetric part of the velocity gradient $\widetilde{\vec{G}}_g$ in \eqref{j_2} would not make any contribution. This magnetic field model would then be redundant with the velocity field model $\widetilde{\vec{v}}$ in \eqref{v_tilde}. On the contrary, one may choose $\widetilde{\vec{j}}_g$ to be independent of $\widetilde{\vec{\omega}}_g$. In figure \ref{fig_mfloc} we show what \eqref{Bmodel2} then typically looks like: qualitatively at least, this model is fairly good, especially given its convenient compact expression, and it is worth confronting to data. However, it may have an important theoretical shortcoming: In \cite{PereiraEtAl16} the authors show that considering two independent white noises in the velocity field \eqref{v_tilde} prevents energy transfer. In our case, we have an additional contribution from shear, so it remains to be seen whether this model has the same limitation or not. A possible improvement may be the following. The two aforementioned cases are two extremes: in one case $\widetilde{\vec{j}}_g$ is fully correlated to $\widetilde{\vec{\omega}}_g$, while it is completely independent in the other. A balance is to consider a $\widetilde{\vec{j}}_g$ `mildly' correlated to $\widetilde{\vec{\omega}}_g$, e.g. a fractional Gaussian vector field such as \eqref{vGaussian}, namely $\widetilde{\vec{j}}_g = \int_{\mathcal{R}_j} \widetilde{\vec{\omega}}_g \times \vec{r} \ \! (r^2+\epsilon_j^2)^{-h_j} \dd V$. On the one hand it has the advantage that we control its statistical properties with simple parameters, and this form guarantees it to be divergence-free\footnote{Which is not required (cf. the similar discussion on $\widetilde{\vec{\omega}}$ at the end of section \ref{sec:buildingDgtilde}), but it is noteworthy, and the impact of considering a divergence-free $\widetilde{\vec{j}}$ or not has to be investigated.}. On the other hand, this introduces yet another set of controlling parameters $\mathcal{R}_j$, $h_j$ and $\epsilon_j$ but which are not physically motivated, unlike the other sets of such parameters.
}

Because \eqref{Bmodel2} is formally similar to the velocity field \eqref{v_tilde}, we may give to the magnetic field $\widetilde{\vec{B}}^{(2)}$ an analogous physical meaning to that presented in the end of section \ref{sec:buildingDgtilde}, replacing vorticity with current density, and adding the effect of shear (contribution from the antisymmetric part of $\widetilde{\vec{G}}_g$, due to local rotation). Hence, \eqref{Bmodel2} introduces an additional regime compared to its hydrodynamical counterpart \eqref{v_tilde}, namely the regime of shear dominated turbulence, when the local rotation $\hat{\vec{\Omega}}$ dominates over vortex stretching $\widetilde{\vec{\mathcal{D}}}_g$.
Compared to model 1, in this model there is no large-scale feature, such as $s_0$ and $\vec{B}_0$. Therefore, in principle this second model is relevant to describe the local, small-scale structure of a turbulent magnetic field. However, given its phenomenological construction, it may turn out to have a wider range of application depending on what it is used for.\\

\textit{Model 3)} Finally, let us propose a model that combines the features of models 1 and 2, by considering again a Biot-Savart law form, but inserting some large-scale feature, \textcolor{Black}{namely
\begin{equation}
\widetilde{\vec{B}}^{(3)} = \frac{\mu_0}{4 \pi} \int_{\mathcal{R}_m} \frac{\widetilde{\vec{j}}^{(3)} \times \vec{r}}{(r^2+\epsilon_m^2)^{h_m}} \dd V,
\label{Bmodel3}
\end{equation}
where
\begin{equation}
\widetilde{\vec{j}}^{(3)} = e^{\tau_m \widetilde{\vec{G}}_g} \ \! \vec{j}_0.
\label{j_3}
\end{equation}
At first sight, this third model looks like the second model, but it is fundamentally different. In $\widetilde{\vec{j}}^{(3)}$ the undeformed current density $\vec{j}_0$ is \textit{not} a random field as in $\widetilde{\vec{j}}^{(2)}$, but an ordered, large-scale current density field. If need be, it may be linked to some ordered magnetic field through Amp\`{e}re's law $\vec{j}_0 = (\vec{\nabla} \times \vec{B}_0)/\mu_0$. In that sense, \eqref{Bmodel3} is somehow `less' random than \eqref{Bmodel2} since the randomness arises only from the matrix exponential. Thus, technically speaking, this third model is a priori more prone to analytical calculations, since the main mathematical difficulty in \eqref{Bmodel2} is that it contains the product of two random distributions \citep[e.g. section 3.4.3 of][]{ChevillardHDR}.}

As in model 2, the Biot-Savart form introduces additional degrees of freedom ($\mathcal{R}_m$, $h_m$ and $\epsilon_m$) compared to model 1, while at the same time, as in model 1, it contains a large-scale field ($\vec{j}_0$). Physically, the magnetic field $\widetilde{\vec{B}}^{(3)}$ is the result of the random stretching/compression and shearing (encoded in the matrix exponential, the strength of which is controlled by $\tau_m$) of a large-scale ordered current density field $\vec{j}_0$, due to the turbulent velocity field \eqref{v_tilde} that it is coupled to.\\

\textcolor{Black}{\cite{PereiraEtAl16} state that the Biot-Savart law form of their hydrodynamical model ensures that their velocity field is divergence-free, as it should since they focus on incompressible dynamics. They also indicate that numerically one has to take a high enough resolution for the size of the grid cells to be small compared to the regularization parameter $\epsilon$. Otherwise gradients are not well approximated and the divergence of the field can be, in standard deviation, of the order of the gradient of one of its components, rather than much smaller.
In our magnetized case, for models 2 and 3 we also use Biot-Savart forms, and we build model 1 from an evolution operator that conserves the divergence-free condition. Indeed, matrix $\vec{G}$ stems from the induction equation \eqref{EvolutionOfB} governing the evolution of magnetic fields, and magnetic fields remain divergence-free throughout time. Hence, the very construction of our models is suited to yield divergence-free fields, as it should. However, a word of caution is in order. We generalize to arbitrary exponents and integration regions these Biot-Savart laws, and we randomize the matrix $\vec{G}$. At this stage we cannot prove rigorously that the magnetic fields indeed remain exactly divergence-free. That being said, we observe numerically that they are divergence-free up to satisfying levels for the range of parameters that we have tested so far, as long as the resolution is high enough (grid cell size greater than the chosen regularization lengths~$\epsilon$), as pointed out in \cite{PereiraEtAl16}. 
In any case, if need be, the simple way out (which is numerically cheap since our formulae are generated in Fourier space) is to apply the projector $\mathcal{P}$ which operates on a given vector $\vec{h}$ as
\begin{equation}
\mathcal{P}(h_i) = \mathcal{F}^{-1} \left[\left(\delta_{ij}-k_i k_j/k^2\right) \mathcal{F}(h_j)\right],
\label{Projector}
\end{equation}
where $\mathcal{F}$ denotes the Fourier transform. Indeed, for any vector $\vec{h}$, $\mathcal{P}(\vec{h})$ is divergence-free. This projector is frequently used in numerical schemes involving magnetic fields, and it also naturally appears in the rewriting (i.e. without approximation) of incompressible MHD \citep[e.g.][]{Galtier16}.
}

We summarize in table \ref{table_DefaultParameters} the many free parameters (and functions) introduced through all the above formulae. These constitute degrees of freedom that may be adjusted to data, in order for example to quantify the degree of intermittency of an observed magnetic field. They also provide a handy tool to model analytically a wide variety of astrophysical environments. Furthermore, this diversity opens a vast parameter space, corresponding to many regimes of the dynamics. For example, since the hydrodynamical and the magnetic regularization parameters $\epsilon_v$ and $\epsilon_m$ correspond to dissipation parameters, it is reasonable to consider that the ratio $\epsilon_v/\epsilon_m$ corresponds to the magnetic Prandtl number $\text{Pr}_m$ (we remain cautious while stating this simply because their precise link with $\text{Pr}_m$ has not been rigorously established yet). Therefore, varying this ratio we may generate turbulent magnetic fields with $\text{Pr}_m \gg 1$ or $\text{Pr}_m \ll 1$, which is out of reach of numerical simulations, as it requires a huge dynamical range. In this sense our approach is complementary to numerical simulations.

\begin{table}
\centering
$
\begin{array}{|c|cc|lllll|}
\cline{2-8}
\multicolumn{1}{c|}{} & \multicolumn{2}{c|}{\text{Expressions}} & \multicolumn{5}{c|}{\text{Degrees of freedom}}\\
\hline
\text{Scalar} & \widetilde{\vec{s}}_g & \eqref{s_gaussian} & \epsilon_s & L_s & h_s &  &  \\
\text{field} & \widetilde{\vec{s}} & \eqref{sf} &  &  &  & \tau_s &  s_0(\vec{x})\\
\hline
\text{Velocity} & \widetilde{\vec{\mathcal{D}}}_g & \eqref{Dg_tilde} & \epsilon_{\omega} & L_{\omega} & h_{\omega} &  &  \\
\text{field} & \widetilde{\vec{v}} & \eqref{v_tilde} & \epsilon_v & L_v & h_v & \tau_\omega & \\
\hline
& \widetilde{\vec{G}}_g & \eqref{Ggtilde} & \epsilon_{\omega} & L_{\omega} & h_{\omega} &  &  \\
\text{Magnetic} & \widetilde{\vec{B}}^{(1)} & \eqref{Bmodel1} &  &  &  &  \tau_m &  \vec{B}_0(\vec{x}) \\
\text{fields} & \widetilde{\vec{B}}^{(2)} & \eqref{Bmodel2} & \epsilon_m & L_m & h_m & \tau_m &  \\
& \widetilde{\vec{B}}^{(3)} & \eqref{Bmodel3} & \epsilon_m & L_m & h_m & \tau_m & \vec{j}_0(\vec{x})\\
\hline
\end{array}
$
\caption{Table of the degrees of freedom (i.e. the free parameters and free functions) appearing in the expressions of the scalar, velocity and magnetic fields introduced in this paper. For simplicity we consider the case of spherical regions $\mathcal{R}_i$ of radius $L_i$, with $i=s,\omega,v$ and $m$.}
\label{table_DefaultParameters}
\end{table}

\section{Examples}
\label{section:Examples}

\textcolor{Black}{The aim of the present work is to provide a tool for observers or anybody who needs to quickly generate synthetic data with controllable properties (statistics and shape of the structures). We do not claim that our formulae constitute solutions of the MHD equations. Rather, we show a way to build non trivial vector fields with many degrees of freedom that are as physically motivated as possible, in order to be intuitive to handle, based on the MHD equations.
It is only in a following paper that we will derive quantitative results on the statistics of the proposed fields, to evaluate how close to, or how far from, real dynamo-generated fields they really are.}

The many degrees of freedom summarized in table \ref{table_DefaultParameters} suggest rich and interesting interplays between the hydrodynamical and magnetic parameters. But an exhaustive analysis of the effects of each of them is also beyond the scope of this paper. Instead, in this section we expose just a couple of selected examples, intending to give the reader a flavor of some key features.

To begin with, let us provide some relevant values for the free parameters, based on \cite{ChevillardEtAl10,ChevillardEtAl11,ChevillardEtAl12,ChevillardHDR}: (i) in the Gaussian scalar field \eqref{s_gaussian} one may take $h_s = 3/4$ typically, (ii) in the works just quoted, it is shown that $h_v=13/12$, corresponding to a Hurst exponent $H=1/3$ in their notation, makes the velocity field \eqref{v_tilde} satisfy the experimental $2/3$ law of turbulence, (iii) in addition, it is argued that in order for the process to have a finite variance and convenient mathematical properties, one should take $3/4<h_v<5/4$ (corresponding to $0<H<1$), a constraint that may thus be taken as a prior in data analyses, (iv) in order for the components of $\widetilde{\vec{\mathcal{D}}}_g$ to be correlated logarithmically in space, $h_{\omega}=7/4$, according to \cite{Mandelbrot72} and \cite{Kahane85}, (v) in the context of laboratory fluids, the value $\tau_\omega = 0.1$ is obtained from experiments, after translating the $\gamma^2=6.7 \times 10^{-2}$ of \cite{PereiraEtAl16} into our notations. Indeed, in laboratory and numerical flows, these authors find a universal behavior, independent of the flow geometry and the Reynolds number, for the longitudinal increments, with data well reproduced for that particular value of $\tau_\omega$.
Finally (vi) regarding the choices of $L_v$ and $\epsilon_v$, in order to reach a large inertial range, one should maximize the $L_v/\epsilon_v$ ratio. The small-scale cut-off $\epsilon_v$ is typically taken equal to a few times the resolution length. For instance for a simulation performed in a 1-periodic box with $N^3$ collocation points, $L_v=1/2$ and $\epsilon_v=3/N$.

We now show examples of various scalar fields, velocity fields and magnetic fields based on the above formulae. Most figures were produced using Mayavi, an application and library for interactive scientific data visualization and 3D plotting in Python \citep{RamachandranVaroquaux11}.

\begin{figure*}
\centering
\includegraphics[scale=0.37]{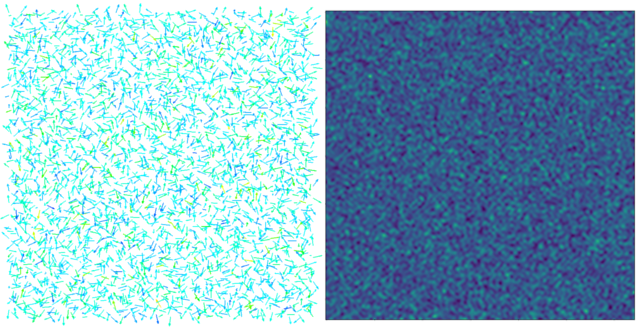}
\caption{All the formulae in this paper originate from a single white noise $\widetilde{\vec{\omega}}_g$. Therefore to generate the scalar fields, velocity fields and magnetic fields of figures \ref{fig_EllipsoidalIntegrationRegion}, \ref{fig_sf}, \ref{fig_vf}, \ref{fig_mfB0}, \ref{fig_mfJ0} and \ref{fig_mfloc} we used the same realization of a Gaussian random vector field, namely the one shown in this figure. On the left panel, we show a 2D slice of this 3D realization, and its norm on the right.}
\label{fig_wn}
\end{figure*}

Firstly, we need to generate the Gaussian vectorial field $\widetilde{\vec{\omega}}_g$, from which all the other fields shown in this paper were generated, making them by construction highly correlated. We display this white noise in figure \ref{fig_wn}. What is interesting to bear in mind when visualizing this realization, is that obviously it does not have any particular shape or structure, neither in vector form nor in norm. $\widetilde{\vec{\omega}}_g$ is really only the building block that gives all the other fields their randomness: the structures and non-Gaussianity are produced by the non-linear transformations \eqref{sf}, \eqref{v_tilde}, \eqref{Bmodel1}, \eqref{Bmodel2}, and \eqref{Bmodel3} applied to it.
Thanks to this, if one varies continuously the free parameters, despite the randomness, the fields also vary continuously: we do not generate new realizations for each value of the free parameters.

Secondly, let us illustrate the effect of the regions of integration, in the case of a scalar field, which is easier to visualize. $\mathcal{R}_s$ determines the large-scale shape, which may depend on position. In figure \ref{fig_EllipsoidalIntegrationRegion}, we consider a simple ellipsoid, and vary its elongation. The shape of this region roughly determines the trend appearing in the spatial distribution: a spherical region yields rather isotropic `clouds', a cigare-like region yields filamentary structures and a pancake-like region generates more sheetlike structures. In the rest of this paper, we will stick to spherical regions, namely spheres of radius $L_i$ centered at position $\vec{x}$:
\begin{equation}
\mathcal{R}_i(\vec{x}) = \left\{\vec{y} \in \mathbb{R}^3 / |\vec{r}| \leq L_i \right\},
\label{SphericalIntegrationRegion}
\end{equation}
where $i=s,v,m$ for respectively a scalar field, a velocity field, and a magnetic field. We do so both for simplicity and because in the literature this has been proven to be relevant at least for the local structure of isotropic turbulence.
	
\begin{figure*}
\centering
\hspace{-0.5cm}
\includegraphics[scale=3.3]{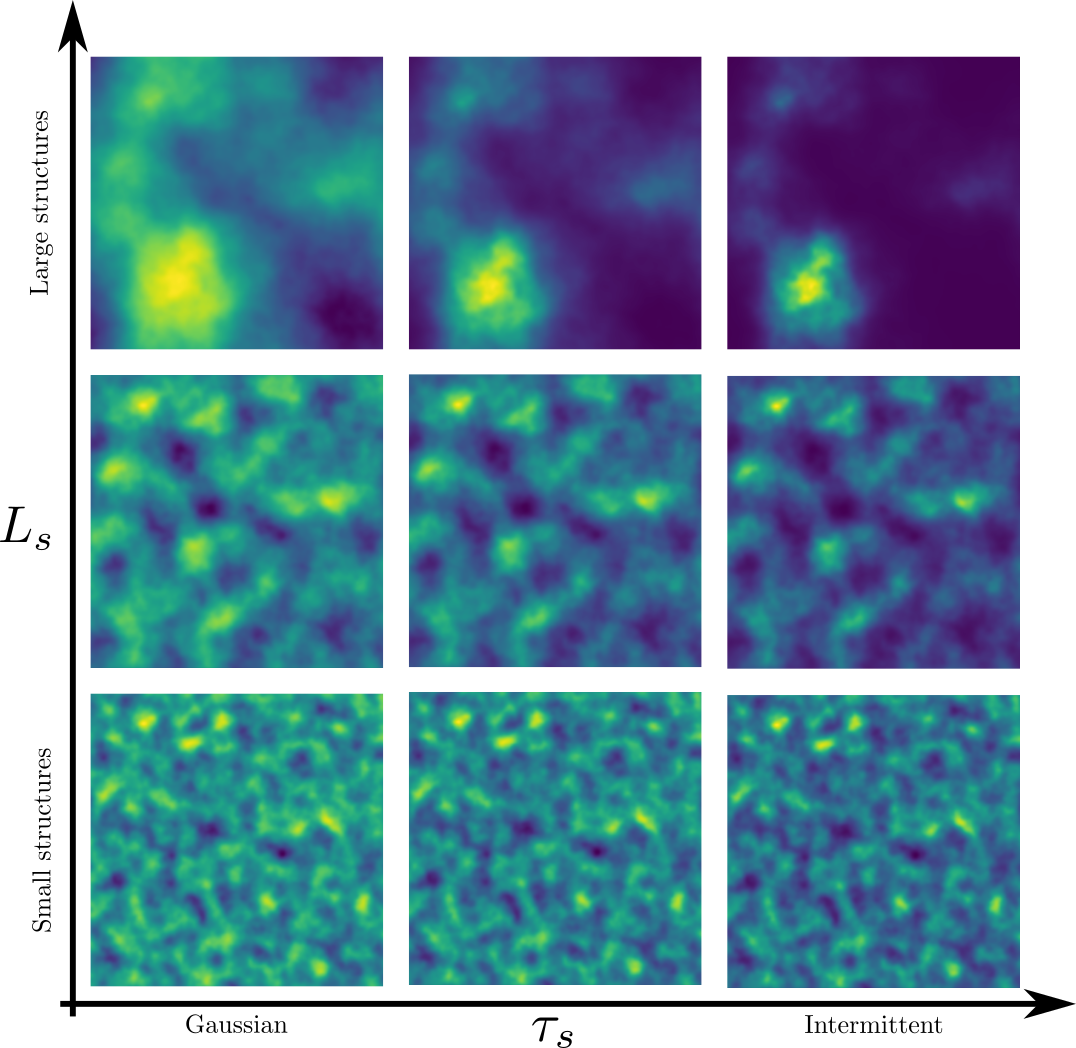}
\caption{This figure shows qualitatively how the scalar field \eqref{sf} varies with the integral scale $L_s$ and the intermittency parameter $\tau_s$. Each panel in this array is a 2D slice of 3D scalar fields given by \eqref{sf}, built using the same white noise realization (namely the one shown in figure \ref{fig_wn}), with the values $L_s=(0.08,0.15,0.5)$ (in fraction of the box size) and $\tau_s=(0.1,0.8,1.6)$. This is the companion figure of figure \ref{fig_pdf_sf}.}
\label{fig_sf}
\end{figure*}

Thirdly, we analyze how the scalar field \eqref{sf} changes as we vary the integral scale $L_s$ together with the intermittency parameter $\tau_s$. In figure \ref{fig_sf}, we show 2D slices of 3D realizations of the scalar field, for various values of $L_s$ and $\tau_s$, in order to show qualitatively how these modify the shape of the structures and the intermittency. Focusing on the modifications along a column, one clearly sees that $L_s$ modifies the global size of the structures formed, i.e. the correlation length. Focusing on the modifications along a row, it appears that basically $\tau_s$ modifies the contrast, and hence controls the intermittency of the field. To complement these intuitive pictures, in figure \ref{fig_pdf_sf} we provide more quantitative considerations by plotting the PDFs of the increments corresponding to each of the cases shown in figure \ref{fig_sf} for three different lags. We normalize the PDFs to unit variance and arbitrarily shift vertically the curves for the sake of clarity \citep[as in][for example]{PereiraEtAl18}. Focusing on a given row, on the left the PDFs are very close to those of a Gaussian random field, while as we move to the right, the characteristic non-Gaussian wings appear as $\tau_s$ increases. This confirms quantitatively the feeling from figure \ref{fig_sf} that $\tau_s$ controls the intermittency of the field. Then considering the modifications along the columns as well, it appears that the `efficiency' of $\tau_s$ to provide intermittency increases with increasing $L_s$.

\begin{figure*}
\centering
\hspace{-0.5cm}
\includegraphics[scale=0.165]{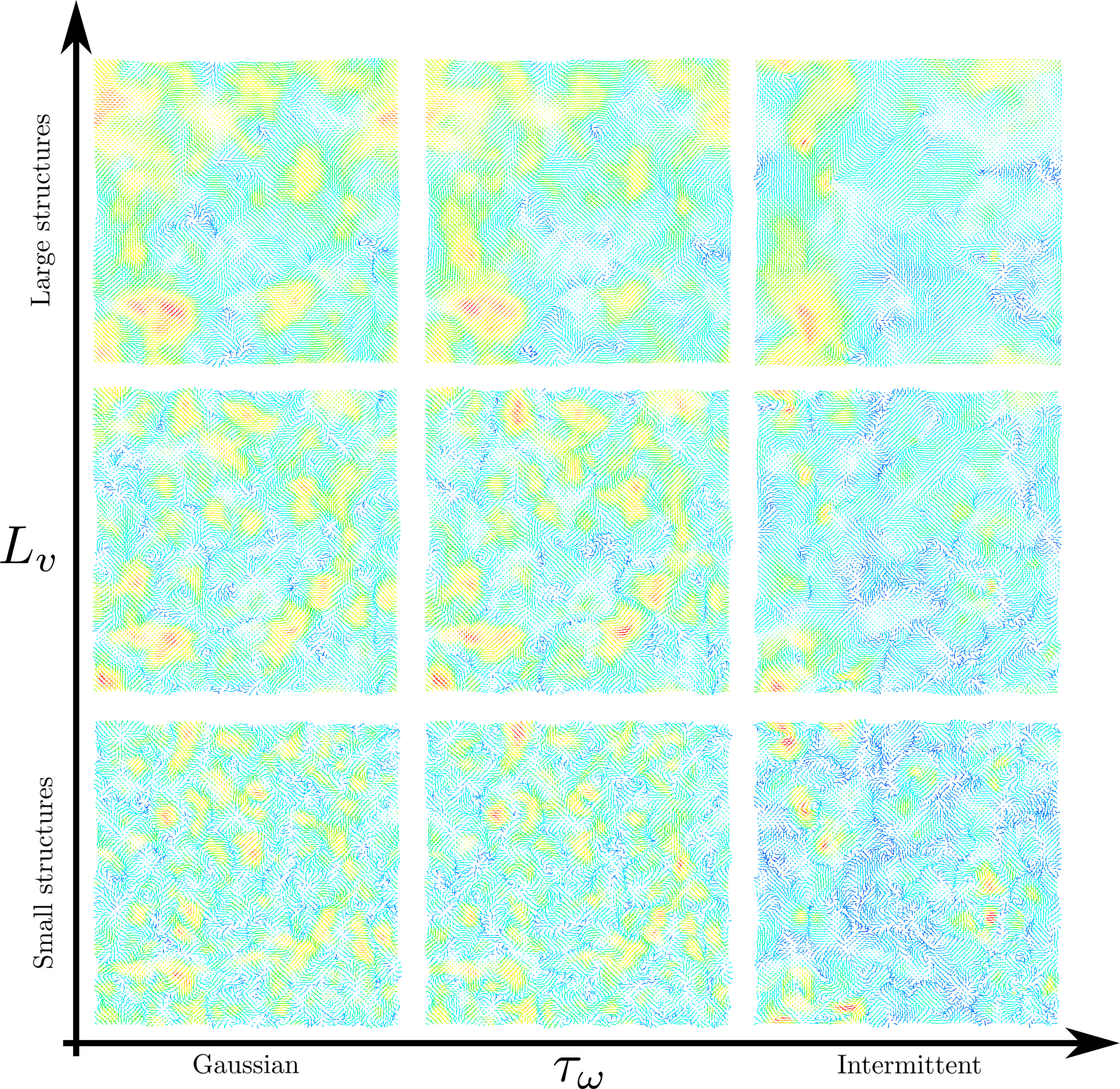}
\caption{This figure shows qualitatively how the velocity field \eqref{v_tilde} varies with the integral scale $L_v$ and the intermittency parameter $\tau_\omega$. Each panel in this array is a 2D slice of 3D velocity fields given by \eqref{v_tilde}, built using the same white noise realization (namely the one shown in figure \ref{fig_wn}), with the values $L_v=(0.08,0.15,0.5)$ (in fraction of the box size) and $\tau_\omega=(0.4,2.5,5)$. Colors indicate the magnitude of vectors, using the light spectrum: red corresponds to the highest velocities and blue to the lowest ones. In 3D all arrows have the same length, so that in these 2D slices the arrows that are small are in fact almost perpendicular to the plane. This way, looking at these 2D representations still gives a feeling of the 3D dynamics. This is the companion figure of figure \ref{fig_pdf_vf}.}
\label{fig_vf}
\end{figure*}

\begin{figure*}
\centering
\hspace{-0.5cm}
\includegraphics[scale=0.21]{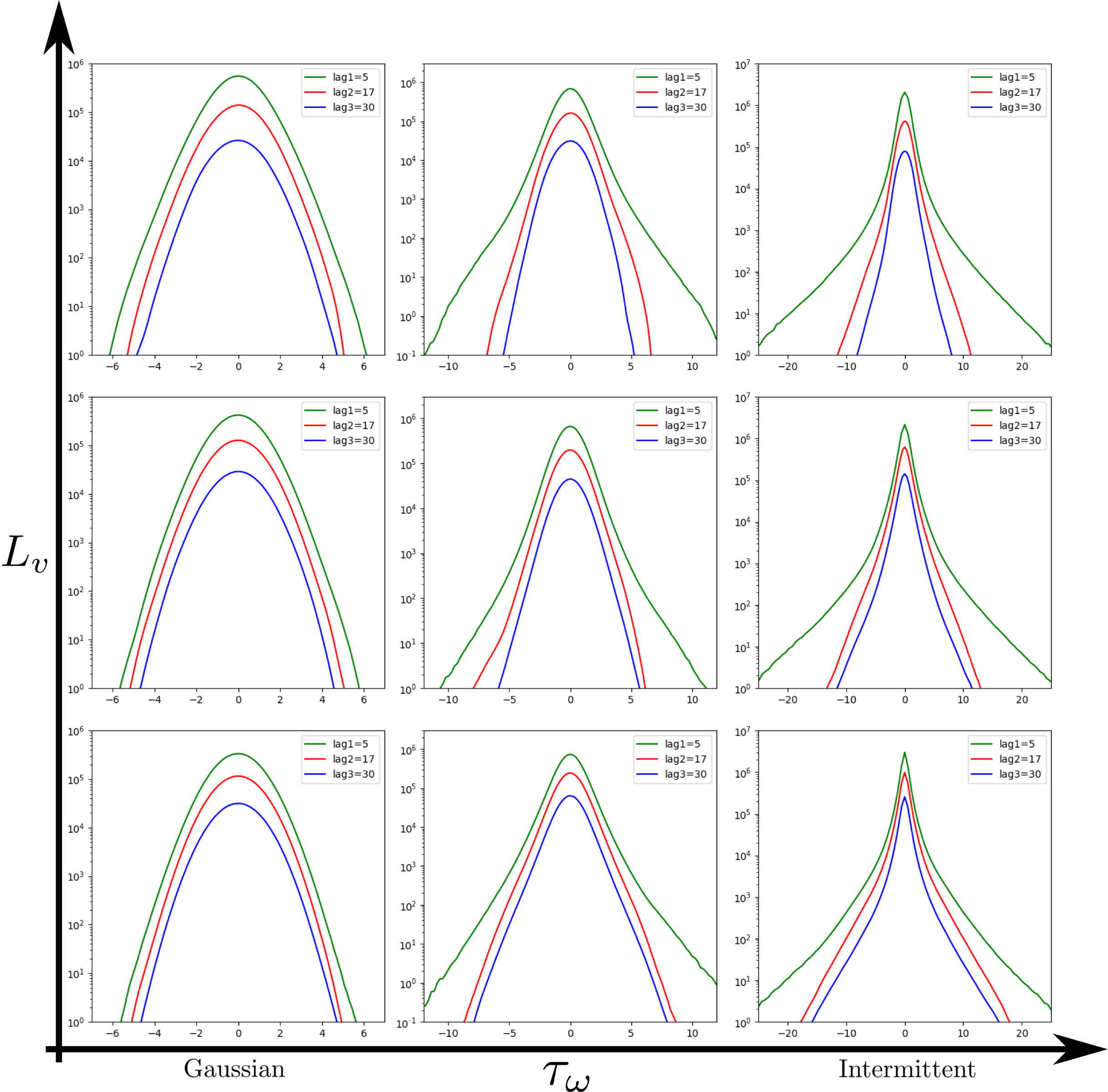}
\caption{PDFs of the longitudinal increments of the velocity field \eqref{v_tilde} for three different lags $\ell = 5,17$ and $30$ (in fraction of the size of the box), averaged over several white noise realizations. The parameters used in the panels of this array correspond to those used in its companion figure, figure \ref{fig_vf}, namely $L_v=(0.08,0.15,0.5)$ (in fraction of the box size) and $\tau_\omega=(0.4,2.5,5)$. This is the companion figure of figure \ref{fig_vf}.}
\label{fig_pdf_vf}
\end{figure*}

\begin{figure*}
\centering
\includegraphics[scale=0.28]{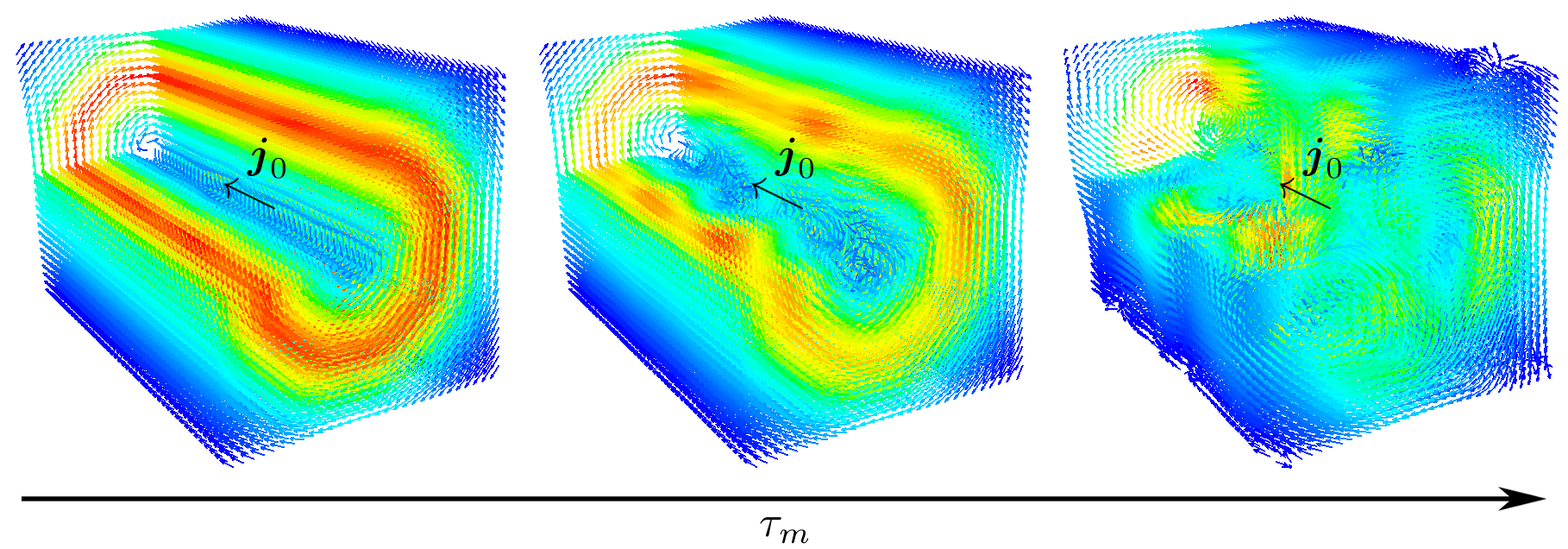}
\caption{3D views of magnetic fields given by our third model \eqref{Bmodel3} using the toy model \eqref{J0_model3} for the ordered current density $\vec{j}_0$, built using the same white noise realization (namely the one shown in figure \ref{fig_wn}), with from left to right the values $\tau_m=(0.01,0.1,0.4)$. The upper left half of each cube is truncated for better visualization, and the choice of representation for the arrows is the same as in figure \ref{fig_vf}. $L_v$ and $L_m$ are both taken equal to a third of the size of the box.
}
\label{fig_mfJ0}
\end{figure*}

Fourthly, we analyze how the velocity field \eqref{v_tilde} changes as we vary the integral scale $L_v$ together with the intermittency parameter $\tau_\omega$. In figure \ref{fig_vf}, we show 2D slices of 3D realizations of the velocity field, for various values of $L_v$ and $\tau_\omega$, in order to show qualitatively how these modify the shape of the structures and the intermittency. Focusing on the modifications along a column, one clearly sees that $L_v$ modifies the global size of the structures formed, i.e. the correlation length. Focusing on the modifications along a row, it appears that basically $\tau_\omega$ modifies the contrast of the norm of the vectors, and hence controls the intermittency of the field.
To complete these intuitive pictures, in figure \ref{fig_pdf_vf} we provide more quantitative considerations by plotting the PDFs of longitudinal increments corresponding to each of the cases shown in figure \ref{fig_vf} for three different lags. As in the scalar field case, non-Gaussian wings appear as we increase the intermittency parameter $\tau_\omega$. These PDFs are very similar to what is usually obtained in direct numerical simulations, but here we may directly control the steepness of the non-Gaussian wings with $\tau_\omega$ and the fields are generated at very little computational cost.
Focusing on a given row, on the left the PDFs are very close to those of a Gaussian random field, while as we move to the right, non-Gaussian wings appear as $\tau_\omega$ increases. This confirms quantitatively the feeling from figure \ref{fig_vf} that $\tau_\omega$ controls the intermittency. Then considering the modifications along the columns as well, it appears that the `efficiency' of $\tau_\omega$ to provide intermittency depends on $L_v$.

\begin{figure*}
\centering
\includegraphics[scale=0.28]{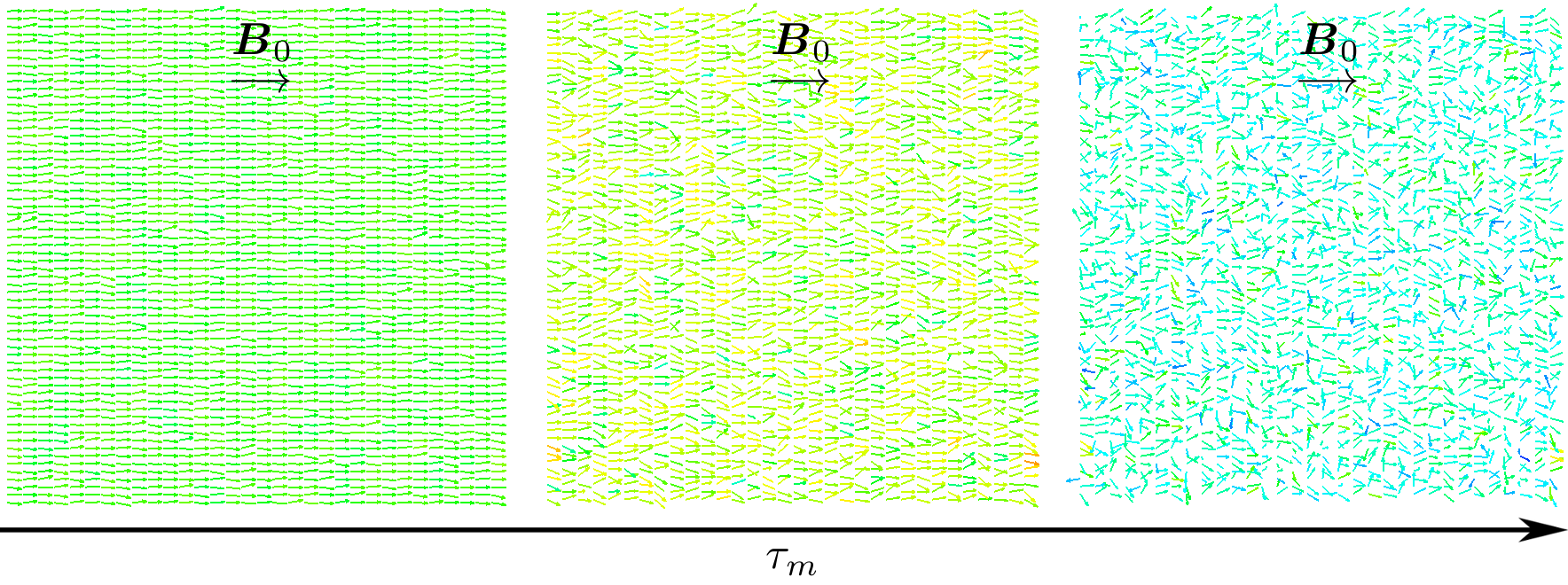}
\caption{These panels show 2D slices of 3D magnetic fields given by our first model \eqref{Bmodel1} using the ordered magnetic field $\vec{B}_0$ model \eqref{B0_model1}, built using the same white noise realization (namely the one shown in figure \ref{fig_wn}), with from left to right the values $\tau_m=(0.2,0.8,1.7)$. The choice of representation for the arrows is the same as in figure \ref{fig_vf}. $L_v$ and $L_m$ are both taken equal to a tenth of the size of the box.}
\label{fig_mfB0}
\end{figure*}

Finally, let us focus on the magnetic fields. To illustrate model 1, we take the following simple divergence-free ordered magnetic field
\begin{equation}
\vec{B}_0=\hat{\vec{x}},
\label{B0_model1}
\end{equation}
where $\hat{\vec{x}}$ is the unit vector along the $x$-direction. This field is then distorted according to \eqref{Bmodel1}, as shown in figure \ref{fig_mfB0} for three values of the magnetic intermittency parameter $\tau_m$. The choice of representation for the arrows is the same as in figure \ref{fig_vf}. In the left panel $\tau_m$ is very small, such that we almost recover the ordered field \eqref{B0_model1}. Then as $\tau_m$ is increased, the vector field is randomized but it is still clear by eye that its statistics are biased, with a tendency of vectors to be pointing to the right, i.e. along the underlying $\vec{B}_0$.

\begin{figure*}
\centering
\includegraphics[scale=0.2]{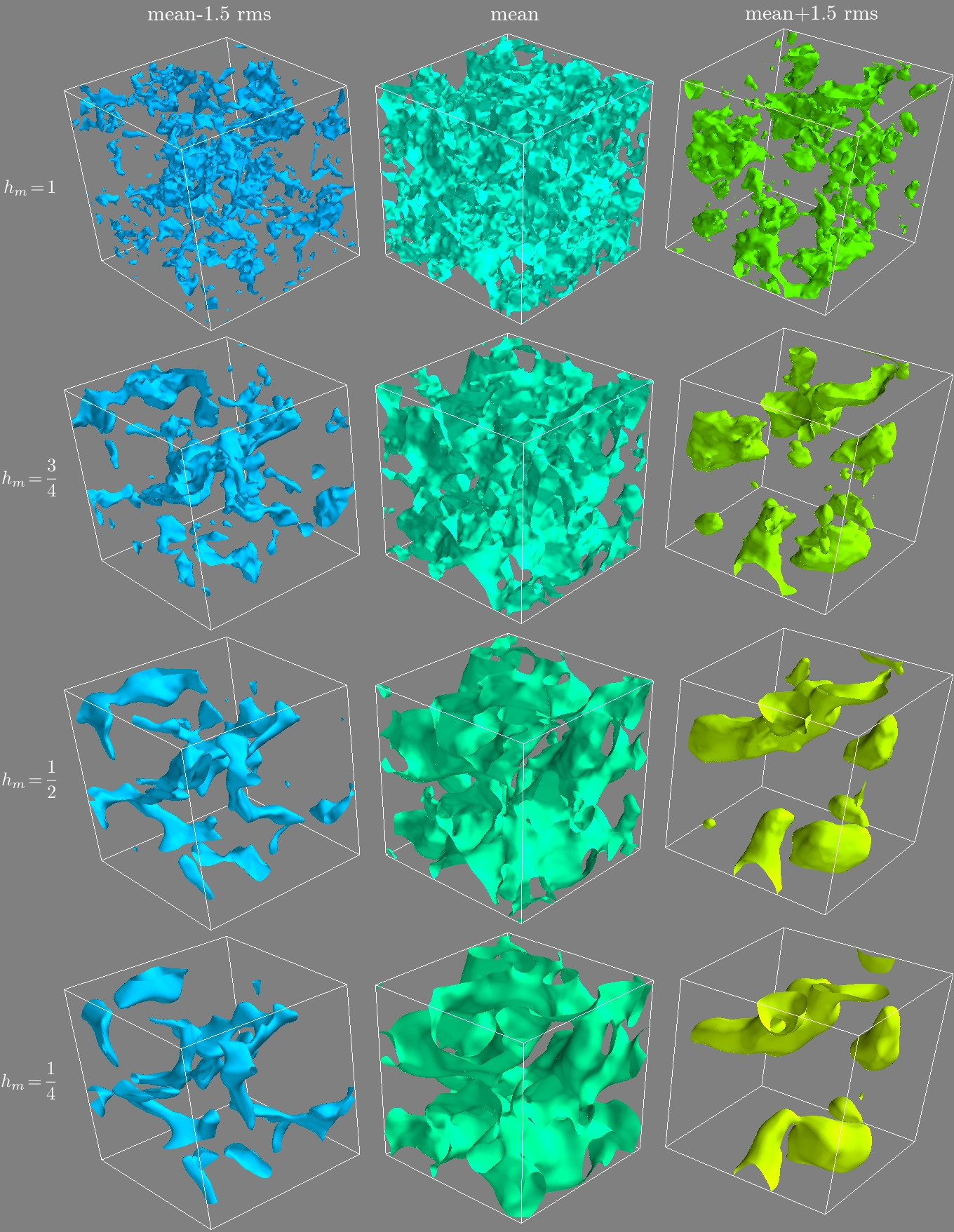}
\caption{
\textcolor{Black}{3D iso-contours of the norm of the magnetic field given by our second model \eqref{Bmodel2}, built using the same white noise realization $\widetilde{\vec{\omega}}_g$ (namely the one shown in figure \ref{fig_wn}), in the case when the Gaussian white noise $\widetilde{\vec{j}}_g$ is independent of $\widetilde{\vec{\omega}}_g$. The left column corresponds to iso-contours in which the norm is equal to the mean minus $1.5$ times the root mean square of the field, in the middle column it is equal to the mean, and in the right it is equal to the mean plus $1.5$ times the root mean square. In these examples we fix $(L_v,\tau_v,L_\omega,\tau_m,L_m)=(0.5,2.5,0.01,2.5,0.5)$, and the Hurst parameters $(h_v,h_\omega)=(13/12,7/4)$, while varying $h_m$: as we go from the top to the bottom row, $h_m$ equals $1,3/4,1/2$ and then $1/4$. As we vary $h_m$ the field varies smoothly between the configurations shown.}
}
\label{fig_mfloc}
\end{figure*}

\textcolor{Black}{To illustrate the second model, we show in figure \ref{fig_mfloc} iso-contours of the norm of the magnetic field \eqref{Bmodel2} in the case when the Gaussian white noise $\widetilde{\vec{j}}_g$ in \eqref{j_2} is independent of $\widetilde{\vec{\omega}}_g$. We do so for four different values of the magnetic Hurst exponent $h_m$, to give a feeling for how this parameter modifies the nature of the field. \textcolor{Black}{Just like the Hurst exponent in fractional Brownian motion is a measure of persistence in a time series, it appears visually here that $h_m$ enables us to tweak the `spatial persistence' of the field, i.e. make it smoother or rougher.} Note that, although the hydrodynamics literature that we quote (which is very rigorous) identifies a range of validity for the Hurst exponents (notably on $h_v$, cf. discussion above), in order to explore the degrees of freedom provided by our formulae, we also considered values beyond these recommended ranges. For example, in the two bottom panels of figure \ref{fig_mfloc}, we use a magnetic Hurst parameter $h_m$ smaller than $3/4$, which does not belong to the range $[3/4,5/4]$ suggested for the hydrodynamical parameter $h_v$ at the beginning of this section. As can be seen in figure \ref{fig_mfloc}, this does result in interestingly looking magnetic fields. In any case, exploring the parameter space freely is even more justified here as we consider the parameters as effective parameters: they are meant to fit astrophysical data, themselves being the result of highly complex dynamics, not captured by our simple initial set of MHD equations, such that ultimately the best fit parameters may not respect the aforementioned priors, which were derived in well-defined, restricted, theoretical frameworks.
}

For the third model, to exemplify the intertwining of the large-scale (ordered) feature with the smaller-scale turbulence, let us consider the following simple case. Obviously we do not claim that it is realistic, but its simplicity helps the visualization. We leave the modeling with more astrophysically relevant ordered fields for future work.
Hence, as an ordered current density field, let us consider a uniform current density wire of radius $r_0$ and norm $j_0$, i.e.
\begin{equation}
\vec{j}_0=
\left\{
\begin{array}{lr}
j_0 \hat{\vec{z}} & \text{ for } r \leq r_0,\\
\vec{0} & \text{ for } r>r_0,
\end{array}
\right.
\label{J0_model3}
\end{equation}
where $r^2 \equiv x^2+y^2$. Pleasingly, in this case the Biot-Savart law \eqref{BiotSavart_Magnetic} can be integrated, so that we know that the corresponding ordered magnetic field is
\begin{equation}
\vec{B}_0= \frac{\mu_0 j_0}{2} (- y \hat{x} + x \hat{y})
\left\{
\begin{array}{cr}
1 & \text{ for } r \leq r_0,\\
\left(r_0/r\right)^2 & \text{ for } r>r_0.
\end{array}
\right.
\label{B0_model3}
\end{equation}
In other words, the vectors $\vec{B}_0$ form concentric circles around the $\hat{\vec{z}}$ axis, with a norm increasing linearly with $r$ below radius $r_0$ and decreasing as $r^{-1}$ beyond $r_0$.
In figure \ref{fig_mfJ0} we show 3D plots of examples of the turbulent magnetic field \eqref{Bmodel3} using the ordered current density \eqref{J0_model3}, for three different values of the magnetic intermittency parameter $\tau_m$. This figure is meant to give a visual impression of how the large-scale features of the ordered magnetic field and the turbulence-related scales are intertwined in this model. On the left, $\tau_m$ is very small, such that we recover the ordered magnetic field \eqref{B0_model3}, as expected in the absence of significant distortion. Then as we increase $\tau_m$, the vectors are more and more randomized. These plots show qualitatively how the statistics of the magnetic field are impacted. The red region in the left panel corresponds to a cylinder of radius $r_0$ (because this is where $\vec{B}_0$ is maximal, cf \eqref{B0_model3}), while in the middle panel for example we can see that the red region is made of lumps. Their typical size basically (in fact there is a non-trivial dependence on $\tau_m$) corresponds to the integral scale $L_m$. Hence we may visualize how the turbulent velocity field distorts the ordered magnetic field $\vec{B}_0$ and modifies its large-scale properties as $\tau_m$ increases. The right panel shows how this information even gets lost at some point once the distortion becomes really strong and one may discern nothing but eddies.

\section{Conclusion $\&$ Prospects}

As an extension to a recent phenomenology of 3D hydrodynamical turbulence based on the so-called multiplicative chaos, we proposed three models for MHD turbulence. These provide compact, physically motivated formulae for a scalar field, a velocity field and three magnetic fields, which are \textit{by construction correlated} to one another, because they all derive from the same white noise. Compared to previous works who focused on the local structure of turbulence, we also introduced large-scale features in these fields, to enlarge their possible range of use. Given the successes exhibited in the hydrodynamics literature concerning the velocity field thus constructed, we have hope that our magnetic field formulae will be useful (i) for data analyses, thanks to the free parameters which provide as many fitting degrees of freedom, (ii) for numerical simulations, providing at low computational cost fairly realistic fields to initialize simulation runs, (iii) as building blocks for analytical toy models of astrophysical environments, and (iv) from a deeper theoretical point of view, the phenomenological steps used here may contribute to improve our understanding of MHD turbulence per se.
We will soon make publicly available a Python code to generate realizations of the fields presented in this paper. Written with the purpose of being user-friendly, it will also include scripts to conveniently analyze and view the 3D fields using Mayavi.

The present work may be improved and extended in numerous ways. In future works, we will analyze both analytically and numerically the statistics of the magnetic fields introduced here, in the line of \cite{PereiraEtAl16}. Then, to better appreciate the scope of these formulae, we will explore more thoroughly the parameter space rather than only a few examples as in section \ref{section:Examples}. It will also be informative to inspect the various regimes of the dynamics. We will in particular look for conditions under which an inverse energy cascade occurs. Earlier, we stressed that it is important to systematically use the same white noise throughout the formulae, because it was shown to guarantee the existence of a direct cascade in the velocity field. Nevertheless, it remains to assess what happens in the magnetized case. Considering different white noises inside the various formulae, the properties of energy transfers through scales will vary depending on the assumed correlations between the noises. Furthermore, from a practical viewpoint for data analysis, this would provide an additional degree of freedom. Another prospect is to compare the properties of our analytical expressions to direct numerical simulations, notably the correlations between the velocity and magnetic fields, \textcolor{Black}{but also regarding the shape of the dissipation regions. Indeed, folded structures (filaments, ribbons, current sheets) naturally form in MHD: In our model considering anisotropic integration regions as well as anisotropic regularization norms (which was taken isotropic here for simplicity) could help generating complex shapes, in addition to the anisotropy introduced by the large-scale ordered fields in the formulae.}

\textcolor{Black}{Another important task will consist in exhibiting relations between the various free parameters (intermittency, Hurst, and regularization) introduced in our models. In the hydrodynamics literature the intermittency parameter $\tau_v$ is also introduced as a free parameter, and to the best of our knowledge, there is no general, simple, result linking it to the other parameters. So far we have found only results derived in special cases, either under simplifying assumptions (one-dimensional fields, asymptotic computations, low order expansions) or valid only for restricted ranges of the parameter space. For example some relations between these parameters and the Reynolds number may be found in \cite{ChevillardMeneveau07,ChevillardHDR,PereiraEtAl16}. Modeling the transition between viscous and inertial-range is also an important step for this task, which is explored in the multifractal formalism in \cite{Meneveau96} as well as in \cite{ChevillardEtAl12}.}

In parallel, it is important to improve our physical understanding of the present phenomenology, and crucial to compare it to others \citep[e.g.][]{GoldreichSridhar95_Part1,GoldreichSridhar95_Part2,PolitanoPouquet95,Nazarenko11,MalaraEtAl16,MalletSchekochihin17}. This may possibly require delving back into the references to the mathematics literature that we mentioned in this paper, to acquire a deeper theoretical knowledge about the random measures we handle here. For instance $\exp(\tau_m \widetilde{\vec{G}}_g) \ \! \dd \vec{W}$ is the product of two random distributions, which is a subtle mathematical object per se.

Finally, after surveying the scope and limitations of our description, one may add more ingredients to improve it. The on-going developments in the hydrodynamics community \citep[e.g.][who study further the dynamics of the velocity gradient]{PereiraEtAl18} remain a wealth of information. Otherwise, so far we have introduced large-scale features by inputting some ordered magnetic and current density fields, but another important large-scale aspect to take into account are the boundaries of the system. A promising starting point to extend our work in that respect is the Biot-Savart operator of a bounded domain, as studied in \cite{EncisoEtAl2018} for example. On the longer term, we will also propose generalizations to compressible turbulence, and go beyond the kinematic regime, to analyse magnetic field dynamos in this formalism. 

\section*{Acknowledgments}

\textcolor{Black}{We express gratitude to the referee for having read our paper so carefully and for his constructive comments. We thank L.~Chevillard for pleasant and fruitful discussions, as well as J.-S.~Carri\`ere, R.~Pereira, B.~Regaldo, T.~Richard, and A.~Seta for help with the coding part of this work.} This research is supported by the Agence Nationale de la Recherche (project BxB: ANR-17-CE31-0022).

\bibliographystyle{mnras}
\bibliography{BxC_theory}

\appendix

\section{Technical aspects}
\label{appendix:TechnicalAspects}

\textcolor{Black}{
In this appendix, we detail some technical points about stochastic calculus, to clarify and justify our choices of notations.}

\textcolor{Black}{
For illustration, let us consider a basic example of a stochastic differential equation, namely a one dimensional It\^o diffusion. In mathematicians' wording, it consists of a time dependent process $X_t$ satisfying
\begin{equation}
\dd X_t = b(X_t) \ \! \dd t + \sigma(X_t) \ \! \dd W_t.
\label{MathViewpoint}
\end{equation}
In the right hand side, the first term is called the drift term with amplitude $b$, and corresponds to the deterministic part. The second term on the contrary is stochastic. It is a noise term with amplitude $\sigma$, where $W_t$ is called a Wiener process, and has the following properties.
}

\textcolor{Black}{
Firstly, while $W_t$ is continuous (almost) everywhere, it is differentiable (almost) nowhere. It is not a regular function, but a distribution. This is why in principle one cannot simply `divide by $\dd t$' the differential relation \eqref{MathViewpoint}, but needs to generalize the notion of differentiation, in the framework of distributions. `Gaussian white noise' corresponds to the (generalized) derivative of the Wiener process. However, it is often convenient to stick to classical notations, and another common way of writing \eqref{MathViewpoint} is
\begin{equation}
\dot{X}=b(X)+\sigma(X) \ \! \eta(t),
\label{PhysicsViewpoint}
\end{equation}
notably in physics \citep[e.g.][]{VandenEijndenBalescu96,BouchetEtAl18}. A historical example in which such a noise term appears is Langevin's equation to model the motion of a particle in a fluid where collisions with the surrounding molecules are modeled as a stochastic force term by means of the white noise $\eta$.
}

\textcolor{Black}{
Secondly, the noise term in \eqref{MathViewpoint} is a random variable, so it should be thought of in terms of its statistical properties. Now, since $W_t$ is a Gaussian process, it is completely specified by its mean and covariance, respectively $\mathbb{E} [W_t] = 0$ and $\mathbb{E} [W_{t'} W_t] = \min(t',t)$, where $\mathbb{E}$ is the expectation value. It can be shown that the covariance can also be specified via $\mathbb{E} [(W_{t'}-W_t)^2] = |t'-t|$. As $t'$ moves closer to $t$, this indicates that $\dd W$ has a variance equal to the differential element $\dd t$. In that sense $\dd W$ behaves as\footnote{This is similar to the fact that in a simple random walk, the typical distance from the origin after $n$ time steps is proportional to $\sqrt{n}$.} $\sqrt{\dd t}$. Hence, in the right hand side of \eqref{MathViewpoint}, while the deterministic term is an infinitesimal element of order $1$, the noise term is an infinitesimal element of order $1/2$. Therefore, in Taylor expansions, the term $\dd W^2$ is of first order, like $\dd t$, and \textit{not} of second order, like $\dd t^2$ (cf. It\^o's lemma). For this reason, rules in stochastic calculus, as basic as the chain rule, differ from `usual' calculus and one has to be cautious with some habits when performing analytical calculations.}

\textcolor{Black}{
In the literature, various notations and terminologies can be found for the above noise term. Firstly, while some call $W_t$ a Wiener process in honor of the mathematician N.Wiener, others refer to it as Brownian motion and write it $B_t$, in honor of the botanist R.Brown.
Secondly, stochastic integrals should in principle be understood in the framework of measure theory, by means of Lebesgue integrals, so mathematically oriented authors call $\dd W$ a Gaussian white measure. Here we adopt a less rigorous approach, just like the Dirac $\delta$ is commonly handled as a regular function in physics, so we stick to a classical, Riemannian, view of integration.
Thirdly, in the literature quoted in the present paper, both notations $\dd \vec{W}(\vec{y})$ and\footnote{The authors use $\dd y$ for the volume element, but we prefer $\dd V$.} $\vec{W}(\dd V)$ can be found for the infinitesimal element inside stochastic integrals. Because integration is performed over 3D space, the variable $\vec{y}$ replaces time $t$ of the above one-dimensional example. These two notations are equivalent, but we may give them different interpretations. The first one, $\dd \vec{W}(\vec{y})$, may be simply interpreted as a field of differential elements with randomly chosen values: to each position $\vec{y}$ in space we associate three values $\dd W_i$ drawn independently following a Gaussian law. The other notation, $\vec{W}(\dd V)$, may be seen as applying the random process $\vec{W}$ to the differential element $\dd V$ at each position $\vec{y}$.
}

\textcolor{Black}{
In the present paper, as physicists, we choose the notation of \eqref{PhysicsViewpoint}, because we consider that overall it makes the method presented in this paper more intuitive and easier to understand. But this is at the cost of introducing an inconvenience in terms of dimensionality. Indeed, for example in \cite{RobertVargas08} the authors, who work in a $d$-dimensional space and have some lengthscale $R$ in their problem, make their initial Gaussian field dimensionless by introducing a factor $R^{d/2}$ in their kernel. It is so because the measure $\dd \vec{W}$ has dimension $R^{d/2}$ in $\mathbb{R}^d$. Hence in fact, to recover this dimensionality, we should replace in our integrals terms like $\widetilde{\vec{\eta}}_g \dd V$ by $\widetilde{\vec{\eta}}_g \sqrt{\dd V}$ (or the more rigorous $\dd \vec{W}$), but it is less customary.
Actually, many authors \citep[e.g.][]{Rodean96} explicitly link the Wiener process to white noise through $\dd W= \eta \ \! \dd t$, rather than $\dd W=\eta \ \! \sqrt{dt}$ as in \cite{ChevillardMeneveau06,ChevillardMeneveau07} for example. In any case, one should bear in mind that these notations have the precise meaning briefly outlined above.
}

\textcolor{Black}{
Numerically, in order to build the Gaussian white noise vector field in a box of length $\ell_B$ with $N$ collocation points in each direction, one may either consider $\vec{W}(\dd V)$ and generate $3 N^3$ independent realizations of a Gaussian variable with mean equal to zero and variance equal to $\dd V=(\ell_B/N)^3$, as in \cite{ChevillardEtAl10,PereiraEtAl16}, or consider $\dd \vec{W} = \widetilde{\vec{\eta}}_g \sqrt{\dd V}$, where $\widetilde{\vec{\eta}}_g$ is a normal distribution with mean equal to zero and variance equal to $1$, as we choose to proceed in this work. For full details on numerical methods for solving stochastic differential equations, see \cite{KloedenPlaten95}.
}

\section{Figures}
\label{appendix:Figures}

\textcolor{Black}{In order not to interrupt the reading of the paper with large figures, we gather all the figures in this appendix.}

\begin{figure*}
\centering
\hspace{-0.5cm}
\includegraphics[scale=0.215]{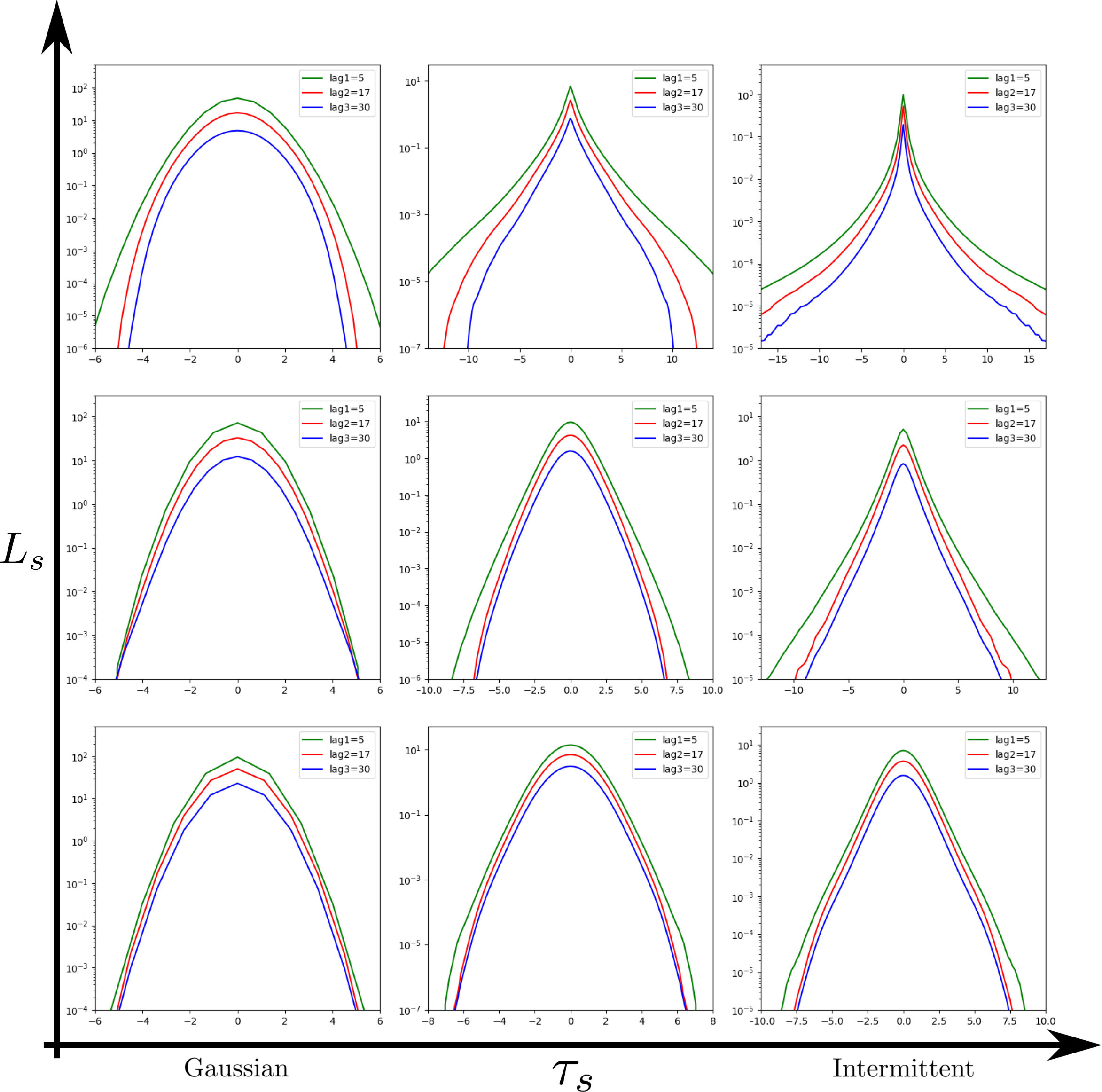}
\caption{PDFs of the increments $\delta_{\vec{\ell}} s(\vec{x}) \equiv s(\vec{x}+\vec{\ell})-s(\vec{x})$ of the scalar field \eqref{sf} for three different lags $\ell = 5,17$ and $30$ (in fraction of the size of the box), averaged over several white noise realizations. The parameters used in the panels of this array correspond to those used in its companion figure, figure \ref{fig_sf}, namely $L_s=(0.08,0.15,0.5)$ (in fraction of the box size) and $\tau_s=(0.1,0.8,1.6)$. This is the companion figure of figure \ref{fig_sf}.}
\label{fig_pdf_sf}
\end{figure*}

\bsp

\label{lastpage}

\end{document}